\documentclass[aps,prd,preprint,superscriptaddress,showpacs,floatfix,nofootinbib,notitlepage]{revtex4-1}
\usepackage{amsmath,graphicx,float,latexsym,hyperref,subfigure,bbold,xcolor,wrapfig}
\usepackage{lineno}
\modulolinenumbers[5]

\newcommand{\jpsi}{\mathrm{J/}\psi}
\newcommand{\psip}{\psi(2\mathrm{S})}

\newcommand{\Wgp}{W_{\gamma\mathrm{p}}}

\newcommand{\Uos}{\Upsilon(\mathrm{1S})}

\begin{document}
\title{Solution to the Balitsky-Kovchegov equation with the collinearly improved kernel including impact-parameter dependence}
\author{D. Bendova}
\author{J. Cepila}
\author{J. G. Contreras}
\author{M. Matas}
\affiliation{Faculty of Nuclear Sciences and Physical Engineering, Czech Technical University in Prague, Czech Republic}
\date{\today}
\begin{abstract}
The solution to the impact-parameter dependent Balitsky-Kovchegov equation with the collinearly improved kernel is studied in detail. The solution does not present the phenomenon of Coulomb tails at large impact parameters that have affected previous studies. The origin of this behaviour is explored numerically. It is found to be linked to the fact that this kernel suppresses large daughter dipoles.
Solutions based on a physics motivated form of the initial condition are used to compute predictions for  structure functions of the proton and  the exclusive photo- and electroproduction of vector mesons. A reasonable agreement is found when comparing to HERA and LHC data. 
\end{abstract}
\maketitle

\section{\label{sec:Intro}Introduction}
Evolution equations are powerful tools to study the  high-energy, equivalently, small-$x$ limit of quantum chromodynamics (QCD)~\cite{Altarelli:1981ax,Lipatov:1996ts,Gribov:1984tu}. The availability of quality data from HERA~\cite{Newman:2013ada} and the LHC~\cite{N.Cartiglia:2015gve} as well as the need for reliable phenomenology for the proposal of new electron-ion facilities~\cite{Accardi:2012qut,AbelleiraFernandez:2012cc} have given an extra impulse to the development of these tools.

In this work, the emphasis is placed on the Balitsky-Kovchegov (BK) evolution equation derived independently in the operator-product-expansion formalism by Balitsky~\cite{Balitsky:1995ub}, and by Kovchegov~\cite{Kovchegov:1999yj,Kovchegov:1999ua} within the colour dipole approach~\cite{Mueller:1989st,Nikolaev:1990ja,Mueller:1993rr}. It corresponds to the large-number-of-colours limit of the Jalilian-Marian-Iancu-McLerran-Weigert-Leonidov-Kovner (JIMWLK) evolution equations~\cite{JalilianMarian:1997gr,JalilianMarian:1997dw,Weigert:2000gi,Iancu:2000hn,Iancu:2001ad,Ferreiro:2001qy}.
The BK equation describes the evolution with rapidity, $Y$, of the dipole-target scattering amplitude, $N(\vec{r},\vec{b}, Y)$, where $\vec{r}$ is the transverse size of the dipole and $\vec{b}$ the impact parameter of the interaction.

Soon after its introduction, the kernel of the leading order BK equation was modified to include corrections that take into account the running of the coupling constant~\cite{Balitsky:2006wa,Kovchegov:2006vj,Albacete:2007yr}. The resulting equation, referred to as rcBK below, 
when combined with appropriate initial conditions -- embodying non-perturbative properties of the hadronic targets -- and disregarding the impact-parameter dependence, produces solutions that have been successfully used to described a wide variety of phenomena.  In particular, the structure function data of the proton as measured at HERA was successfully described~\cite{Albacete:2004gw,Albacete:2009fh,Albacete:2010sy,Cepila:2015qea}. A few other  applications of these solutions are, for example, gluon production in heavy-ion collisions~\cite{ALbacete:2010ad}, single particle~\cite{Lappi:2013zma} and $\jpsi$ production in pp and pA collisions~\cite{Ducloue:2015gfa} , di-hadron correlations in p--Pb interactions~\cite{Albacete:2018ruq} and even the flux of atmospheric neutrinos~\cite{Albacete:2015zra,Bhattacharya:2016jce}.

As already mentioned, these comparisons of rcBK-based predictions to data disregarded the impact-parameter dependence of the  dipole amplitude. The reason is that earlier studies of solutions including the impact parameter found that the amplitude developed a power-like dependence on $b\equiv|\vec{b}|$, the so-called Coulomb tails, which generate an unphysical growth of the cross section~\cite{GolecBiernat:2003ym}. Nonetheless attempts were made to modify the kernel to solve this problem, for example, by adding  an ad-hoc cut-off for large sizes of the daughter dipoles~\cite{Berger:2010sh}. The solutions found had no more Coulomb tails, but needed an extra, so-called {\it soft}, contribution to be able to describe HERA data on structure functions~\cite{Berger:2011ew}. (A similar conclusion also holds for the solutions of the impact-parameter dependent JIMWLK equation~\cite{Mantysaari:2018zdd}.) Nonetheless, this approach did a good job when confronted with HERA data on exclusive vector meson production~\cite{Berger:2012wx}.

Recently, the kernel of the leading order equation has been improved by including the resummation of all double collinear logarithms~\cite{Iancu:2015vea} as well as two classes of single logarithmic corrections~\cite{Iancu:2015joa}. Using this kernel and disregarding the dependence on the impact parameter, it was also possible to obtain a good description of HERA data on the structure function of the proton. Finally, in the rapid communication~\cite{Cepila:2018faq}, we have demonstrated that solutions of the BK equation with the collinearly improved kernel and an appropriate initial condition describe correctly the HERA data on structure functions and the $t$ dependence of the exclusive photoproduction of $\jpsi$ at one energy without the need of any additional ad-hoc parameter or correction. 

In this contribution the studies reported in~\cite{Cepila:2018faq} are extended to discuss in depth the behaviour of the collinearly improved kernel and of the solutions of the corresponding BK equation, comparing them to the rcBK case. In addition, more details on the  comparison to HERA structure function data are presented, and comparison of our predictions to relevant HERA and LHC data on exclusive vector meson photo- and electro-production is provided. In all cases, the agreement between model and measurements is satisfactory.

The rest of this contribution is organised as follows: In Sec.~\ref{review}  the formalism used throughout this work is reviewed. In Sec.~\ref{Solving}  the technical details to solve the collinearly improved impact-parameter dependent BK equation  are addressed. In Sec.~\ref{Evolution} the origin of the  suppression at large impact parameters is discussed, the behaviour of the solution is contrasted  with solutions of the rcBK case, and the shape of the amplitude is shown at different values of rapidity, dipole size and impact parameter. In Sec.~\ref{Data} and  Sec.~\ref{VM}  our predictions are confronted with structure function  data measured at HERA, and to  data for cross sections of exclusive photo- and electroproduction of $\phi$, $\jpsi$, $\psip$, and $\Uos$ vector mesons measured both at HERA and at the LHC, respectively. Sec.~\ref{sec:Conclusions} contains a brief summary of our findings and presents our conclusions.

\section{Review of the formalism}
\label{review}

\subsection{The Balitsky-Kovchegov equation}
\label{BK}
The  BK evolution equation reads~\cite{Balitsky:2006wa,Kovchegov:2006vj}
\begin{eqnarray}\label{fullbalitsky}
\frac{\partial N(\vec{r}, \vec{b}, Y)}{\partial Y} = \int d\vec{r_{1}}K(r,r_{1},r_{2})& \Big(& N(\vec{r_{1}}, \vec{b_1}, Y) + N(\vec{r_{2}}, \vec{b_2}, Y) - N(\vec{r}, \vec{b}, Y) \nonumber \\
& & - N(\vec{r_{1}}, \vec{b_1}, Y)N(\vec{r_{2}}, \vec{b_2}, Y)\Big),
\end{eqnarray}
where $r\equiv|\vec{r}\,|$, $r_1\equiv|\vec{r_1}|$, and $r_2\equiv|\vec{r_2}|\equiv|\vec{r}-\vec{r_1}|$ are the sizes of the original dipole and of the two daughter dipoles, respectively. Note that these are 2-dimensional vectors in the same plane as the impact parameter. The  magnitudes of the corresponding impact parameters are $b\equiv\vec{b}$, $b_1\equiv\vec{b_1}$, $b_2\equiv\vec{b_2}$. The kernel $K(r,r_{1},r_{2})$ is discussed below. 

In this work, the solution to the BK equation is obtained under the  assumption that the scattering amplitude $N(\vec{r}, \vec{b}, Y)$ depends solely on the sizes of the dipoles and of the impact parameter vectors. In practice, this means to solve the equation
\begin{eqnarray}
\frac{\partial N(r,b, Y)}{\partial Y} = \int d\vec{r_{1}}K(r,r_{1},r_{2}) & \Big( & N(r_1,b_1, Y) + N(r_2,b_2 Y) - N(r,b, Y) \nonumber \\
& & - N(r_1,b_1, Y)N(r_2,b_2, Y) \Big), \label{BKused}
\end{eqnarray}
subjected to the condition that the angle between $\vec{r}$ and $\vec{b}$ is fixed. We chose to fix this angle at zero, meaning that these vectors are parallel. 

\subsection{Kernels of the Balitsky-Kovchegov equation}
\label{Kernels}

Several functional forms for the kernel of the BK equation have been proposed. The ones that are mentioned in this work are presented in the following.

The leading order kernel is given by
\begin{equation}\label{LOkernel}
    K_{\rm LO}(r, r_1, r_2) = \frac{\alpha^{\rm nr}_s}{2\pi}\frac{r^{2}}{r_{1}^{2}r_{2}^{2}},
\end{equation}
where the non-running coupling,  $\alpha^{\rm nr}_s$, is fixed to a constant value. 

The running coupling kernel $K_{{\rm rc}}(r,r_{1},r_{2})$ reads~\cite{Balitsky:2006wa}
\begin{equation}\label{Krun}
K_{{\rm rc}}(r,r_{1},r_{2}) = \frac{N_{c}\alpha_{s}(r^{2})}{2\pi^{2}} \left( \frac{r^{2}}{r_{1}^{2}r_{2}^{2}} + \frac{1}{r_{1}^{2}}\left( \frac{\alpha_{s}(r_{1}^{2})}{\alpha_{s}(r_{2}^{2})}- 1\right) + \frac{1}{r_{2}^{2}}\left( \frac{\alpha_{s}(r_{2}^{2})}{\alpha_{s}(r_{1}^{2})}- 1\right) \right),
\end{equation}
where $N_c$ is the number of colours and $\alpha_{s}$ is the running coupling, which is further discussed  in Sec.~\ref{Coupling}.  

The running coupling kernel with a cutoff to tame the Coulomb tails generated by  the evolution in the impact parameter is given by~\cite{Berger:2011ew}
\begin{equation}\label{fullbalitskyredcut}
K_{{\rm rc}}^{\rm bdep}(r,r_{1},r_{2}) = K_{{\rm rc}}(r,r_{1},r_{2}) \Theta\left(\frac{1}{m^2} - r_1^2 \right)\Theta\left(\frac{1}{m^2} - r_2^2 \right),
\end{equation}
where $\Theta$ is the Heaviside function and $m$ a parameter to limit the size of daughter dipoles.

Finally, the collinearly improved kernel is~\cite{Iancu:2015joa}
\begin{equation}\label{collinearlyimproved}
K_{\rm ci}(r, r_1, r_2) = \frac{\overline{\alpha}_s}{2\pi}\frac{r^{2}}{r_{1}^{2}r_{2}^{2}} \left[\frac{r^{2}}{\min(r_{1}^{2}, r_{2}^{2})}\right]^{\pm \overline{\alpha}_sA_1} K_{\rm DLA}(\sqrt{L_{r_1r}L_{r_2r}}),
\end{equation}
where
\begin{equation}
K_{\rm DLA}(\rho) = \frac{J_1(2\sqrt{\overline{\alpha}_s \rho^2})}{\sqrt{\overline{\alpha}_s \rho}},
\label{eq:DLA}
\end{equation}
$J_1$ is the Bessel function (the inclusion of the Bessel function into the BK kernel has been previously discussed in~\cite{Motyka:2009gi}), the anomalous dimension is $A_1= 11/12$, and 
\begin{equation}
L_{r_ir} = \ln\left( \frac{r_i^2}{r^2} \right).
\end{equation}
The sign factor in the exponent $\pm \overline{\alpha}_sA_1$ takes the value of the plus sign when
$r^2 < \min(r_1^2, r_2^2)$ and the negative sign otherwise. For the running coupling
\begin{equation}\label{baralpha}
\overline{\alpha}_s = \alpha_s\frac{N_c}{\pi},
\end{equation}
the smallest dipole prescription is used throughout the computation according to
\begin{equation}\label{minimalapproach}
\alpha_s = \alpha_s (r_{\min}),
\end{equation}
where $r_{\min} = \min(r_1,r_2,r)$.
 This prescription was compared to other prescriptions in~\cite{Iancu:2015joa}, where it was found to work adequately in this context.  This prescription has also been suggested as the natural option for the BK equation at next-to-leading order~\cite{Balitsky:2008zza}. 

\subsection{Treatment of the coupling constant}
\label{Coupling}
In this work the running coupling is computed in the variable-number-of-flavours scheme, implemented according to 
\begin{equation}\label{alph}
\alpha_{s, n_{f}} (r^{2}) = \frac{4\pi}{\beta_{0,n_{f}}\ln\left(\frac{4C^{2}}{r^{2}\Lambda ^{2}_{n_{f}}}\right)},
\end{equation}
where $n_{f}$ corresponds to the number of flavours that are active, $C^{2}$ is an infrared regulator that takes into account the approximations made for the computation of the Fourier transform into the position space and is usually fit to data~\cite{Albacete:2009fh}. The variable $\beta_{0, n_f}$ is the leading order coefficient of the QCD beta-series and is given by relation
\begin{equation}\label{beta}
\beta_{0, n_{f}} = 11 - \frac{2}{3}n_{f}.
\end{equation} 
The value of the QCD scale parameter $\Lambda^{2}_{n_{f}}$ depends on the  number of active flavours. When heavier quarks  are active (charm and beauty quarks), its value is obtained from the relation~\cite{Albacete:2010sy}
\begin{equation}
\Lambda_{n_{f}-1}=(m_{f})^{1-\frac{\beta_{0,n_{f}}}{\beta_{0,n_{f}-1}}}(\Lambda_{n_{f}})^{\frac{\beta_{0,n_{f}}}{\beta_{0,n_{f}-1}}}.	
\end{equation}
This recursive relation needs to be fixed at one point and for this the usual choice is to take the value of the running coupling at the scale of the mass of the Z$^{0}$ boson. In this way, $\Lambda_{5}$ is set with the use of the experimentally measured value of $\alpha_{s}(M_{Z})=0.1196 \pm 0.0017$, where the Z$^{0}$ mass is $M_{Z}=91.18$\,GeV$/c^2$~\cite{Agashe:2014kda}. 
The number of active flavours  is set depending on the transverse size of the mother dipole. The condition that governs this relates the mass of the heaviest quark considered to the values of the dipole size $r$. This condition can be expressed as
\begin{equation}
r^{2} < \frac{4C^{2}}{m_{f}^{2}}.
\end{equation}\label{active}
Since all dipole sizes are accounted for in the BK evolution equation, there is a need to freeze the coupling at a set value after a certain dipole size is reached ~\cite{Albacete:2009fh}. In this work, the coupling is frozen at  $\alpha_s^{\rm sat}$ = 1 as in~\cite{Iancu:2015vea}.

\begin{figure}[!ht]
  \centering
  \includegraphics[width=0.75\linewidth]{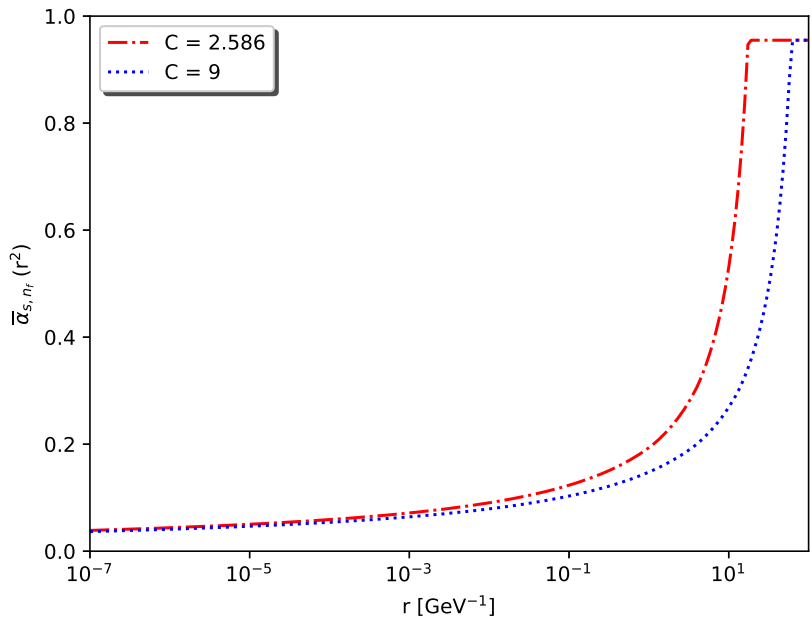}
\caption{(Colour online) Comparison between the behaviour of $\overline{\alpha}_s$ computed from Eqs.~(\ref{baralpha}) and~(\ref{alph}) with $C$ = 2.586 (red) and C = 9 (blue).}
\label{alphacomparison}
\end{figure}

The value of the parameter $C$ affects the description of data by modifying the speed of the evolution and effectively changes the slope of the structure function. The higher value of this parameter the more   the running of the  coupling is suppressed and, consequently, the slope in the structure function $F_2$ is less steep. Figure~\ref{alphacomparison} compares the running of  $\alpha_s$ for two values of $C$: the one used here, $C=9$, and the one used in~\cite{Iancu:2015vea}, $C=2.586$. The value $C=9$ was set heuristically and since the solutions reproduce correctly the data, as shown below,  it has  not been further optimised.

\section{Impact-parameter solution to the Balitsky-Kovchegov equation}
\label{Solving}

\subsection{Initial condition}
\label{Initial}
\begin{figure}[ht]
  \centering
  \includegraphics[width=0.4\linewidth]{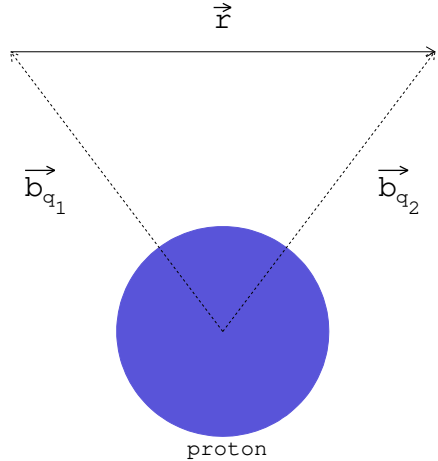}
\caption{Schematic picture of the variables that enter the initial condition presented in Eq.~(\ref{initialeq}).}
\label{initialcondition}
\end{figure}

The initial condition, already introduced in~\cite{Cepila:2018faq}, depends on the impact parameter; it is suppressed in the regions where  $r$ or $b$ reaches large values, in order to respect  the geometric nature of the dipole-proton interaction. The shape of its functional form is a combination of the expected behaviour in $r$, which is obtained from the Golec-Biernat W\"usthoff (GBW) model~\cite{GolecBiernat:1998js}, and the impact-parameter dependence, which uses a Gaussian distribution to reflect the expected  profile of the proton. Such an approach has been used in similar forms in the past; e.g., in~\cite{Kowalski:2003hm,Watt:2007nr,Marquet:2007nf,Mantysaari:2016ykx,Cepila:2016uku}. The main new ingredient with respect to the initial condition used in the previous studies~\cite{Berger:2010sh,Berger:2011ew,Berger:2012wx} is the explicit separation of  the contribution from the individual quark and anti-quark forming the dipole. The initial condition is given by 
\begin{equation}\label{initialeq}
N(r, b,Y=0) =  1 - \exp\left(-\frac{1}{2}\frac{Q^2_s}{4}r^2 T(b_{q_1},b_{q_2})\right),
\end{equation}
where $b_{q_i}$ are the impact parameters of the quark and anti-quark forming the dipole and
\begin{equation}
T(b_{q_1},b_{q_2})= \left[\exp\left(-\frac{b_{q_1}^2}{2B_G}\right) + \exp\left(-\frac{b_{q_2}^2}{2B_G}\right)\right].
\end{equation}
As a first attempt,  the angle between $\vec{r}$ and $\vec{b}$ was fixed  as shown schematically  in Fig.~\ref{initialcondition}. As the results obtained with this initial condition are satisfactory, no further optimisation has been considered.

The parameters appearing in this initial condition, $Q^2_s$ and $B_G$, have a clear physical interpretation as the saturation scale and as the variance of the Gaussian distribution of the target in impact parameter, respectively. The value of the $Q^2_s$ parameter
 is chosen to be 0.496 GeV$^2$, such that the $F_2(x, Q^2)$ data are correctly described at the rapidity of the initial condition. The relation between $x$ and rapidity is $Y = \ln\big(x_0/x\big)$, where $x_0=0.008$. The parameter $B_G$ is set to 3.2258 GeV$^{-2}$  in order to describe the data for exclusive photoproduction  of $\mathrm{J/}\psi$ off protons at a photon--proton centre-of-mass energy $\langle W_{\gamma p} \rangle = 100$ GeV, where, as customary $x=(M^2+Q^2)/(W_{\gamma p}^2+Q^2)$ is used; here,  $M$ represents the mass of the vector meson.
 
\subsection{Setup for the numerical solution to the equation}
\label{Numerics}
The BK evolution equation does not have an analytic solution and therefore has to be solved numerically. The procedure used by us in~\cite{Cepila:2015qea,Matas:2016bwc} was extended to the case of the impact-parameter dependent BK equation~\cite{Cepila:2018faq}  and the solution is evolved in rapidity with a step of $\Delta Y$ = 0.01. 

Fixed grids are used for $r$ and $b$. They  are logarithmic grids of base 10 with  25 evenly-spaced points per order of magnitude, spanning the range from 10$^{-7}$ to 10$^4$ GeV$^{-1}$ for both the $r$ and $b$ variables. The integration over $\vec{r_1}$ is performed in polar coordinates, where $r_1$ is evaluated in the same grid as $r$ and the polar angle, denoted by $\theta_{rr_1}$, is evaluated in a fixed grid with 21 points separated by a constant step. The numerical integrations are performed applying  Simpson's method.

Since the transverse dipole vectors are related as $\vec{r} = \vec{r_1} + \vec{r_2}$, by fixing the values of $r$ and $r_1$ to the pre-defined grid,  the values of $r_2$ are in general off the grid. Whenever this happens, linear interpolation in the log$_{10}$ space is used to get the desired value of $N(r_2, b_2, Y)$. A similar approach is used for obtaining the value of the scattering amplitude whenever the value of $b_1$ or $b_2$ is off the  grid. 

The values of $b_1$ and $b_2$ are then computed from the relations $\vec{b_1} = \vec{b} + \vec{r_2}/2$ and $\vec{b_2} = \vec{b} - \vec{r_1}/2$ assuming a fixed angle between $\vec{r}$ and $\vec{b}$. As mentioned above, this angle is set to zero for the results presented below.

The solution to the BK equation has been implemented independently using C++ and the Intel Fortran Compiler. Both implementations have similar performance, with the Fortran version being slightly faster.
In a standard personal computer, the program performs the evolution of the dipole  amplitude in one unit of rapidity, that is 100 steps for the settings described above, in a bit less than one hour for one set of parameters. 

To test the numerical stability of the selection of the grid, the setup was modified and the scattering amplitude was compared at $Y=3$, $r=1$/GeV and all values of $b$. We have changed the step in rapidity from  0.01 to 0.02, the number of steps in $r$ and $b$ per order of magnitude from 25 to 15 and the size of the grid in the polar angle from 21 to 16 and 31 points. Except for the change to 16 points in the grid for polar angles, all other changes produced a difference below the per-mil level. The use of the spare grid in polar angle produced changes almost at one percent level.  In summary, with the chosen settings a numerical precision at the percent level, or even below it, is expected.

\section{The solution to the BK equation}
\label{Evolution}

\subsection{Behaviour of the collinearly improved kernel}
\label{ciKernel}

\begin{figure}[!ht]
  \centering
  \includegraphics[width=0.65\linewidth]{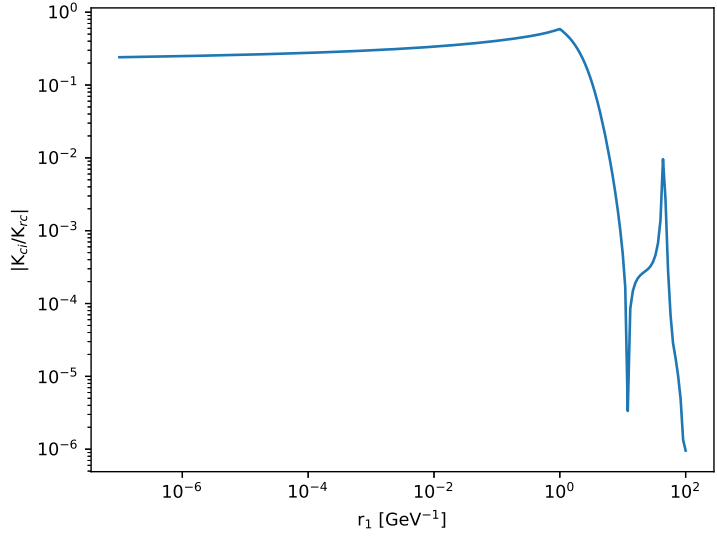}
\caption{Absolute value of the ratio $K_{\rm ci}/K_{\rm rc}$ at a fixed dipole size $r=1\,\mathrm{GeV}^{-1}$ and orientation with respect to the daughter dipole $\theta_{rr_1}=\pi/2$ as a function of the daughter dipole size.}
\label{kernel_ratio}
\end{figure}

As was shown in~\cite{Cepila:2018faq}, the solutions to the BK equation do not exhibit Coulomb tails when using the collinearly improved kernel. This  behaviour  is related to the suppression of this kernel for large values of the size of the daughter dipoles. As an illustration, Fig.~\ref{kernel_ratio} shows the ratio of the collinearly improved kernel, see Eq.~(\ref{Krun}), to the running-coupling kernel, see Eq.~(\ref{collinearlyimproved}). (The parameter $C$ for the running coupling in this kernel was chosen to be $C=9$ just as in the collinearly improved kernel for the sake of a valid comparison.)
The ratio is computed at $r$ = 1\,GeV$^{-1}$ and $\theta_{rr_1}=\pi/2$. Other values produce a similar picture. The figure shows that for large sizes of the daughter dipole the collinearly improved kernel is orders of magnitude smaller than the running-coupling one. 

To follow up in more detail the origin of this behaviour the kernels are divided into three parts. For the collinearly improved kernel, they are
\begin{align}
& K_{\rm ci}^1 = \frac{\overline{\alpha}_s}{2\pi}\frac{r^{2}}{r_{1}^{2}r_{2}^{2}}, \\
& K_{\rm ci}^2 = \left[\frac{r^{2}}{\min(r_{1}^{2}, r_{2}^{2})}\right]^{\pm \overline{\alpha}_sA_1}, \\
& K_{\rm ci}^3 = K_{\rm DLA}(\sqrt{L_{r_1r}L_{r_2r}}).
\end{align}
The first term, $K_{\rm ci}^1$, is present already at the leading order if one considers a fixed value of the running coupling, $K_{\rm ci}^2$ takes into account the contribution from the single collinear logarithms, and $K_{\rm ci}^3$ resumms double collinear logarithms to all orders. The entire collinearly improved kernel is then given by the multiplication of all these factors as
\begin{equation}
    K_{\rm ci} = K_{\rm ci}^1 K_{\rm ci}^2 K_{\rm ci}^3.
\end{equation}

\begin{figure}[!ht]
  \centering
  \includegraphics[width=0.49\linewidth]{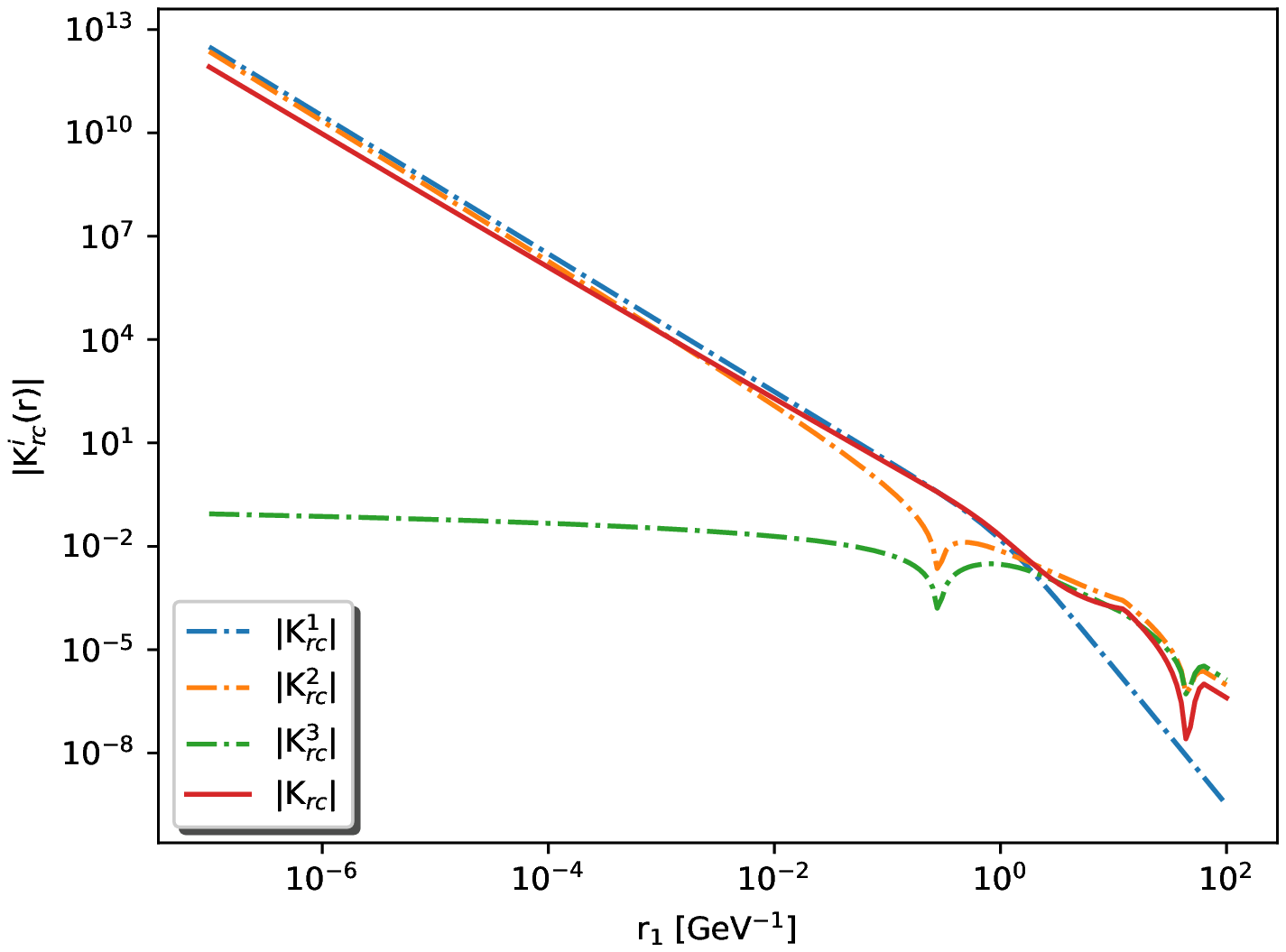}
  \includegraphics[width=0.49\linewidth]{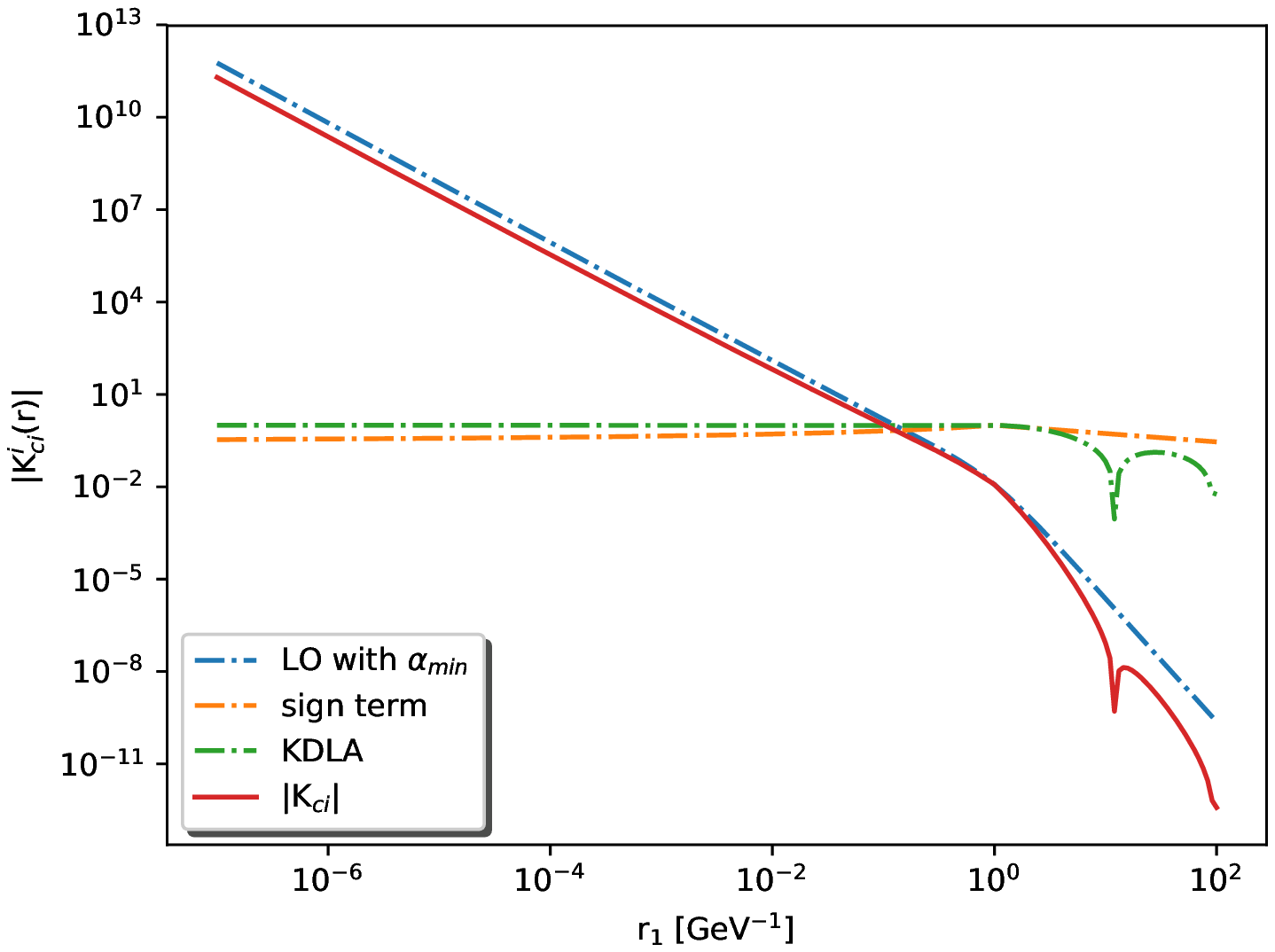}
\caption{The three constituent terms of the BK kernel for the running coupling (left) and collinearly improved cases (right) at a fixed dipole size $r=1\,\mathrm{GeV}^{-1}$ and orientation with respect to the daughter dipole $\theta_{rr_1}=\pi/2$.}
\label{kernel_elements}
\end{figure}

For the running coupling BK kernel, the separation in three parts is as follows:
\begin{align}
& K_{\rm rc}^1 = \frac{N_{c}\alpha_{s}(r^{2})}{2\pi^{2}}\frac{r^{2}}{r_{1}^{2}r_{2}^{2}},\\
& K_{\rm rc}^2 = \frac{N_{c}\alpha_{s}(r^{2})}{2\pi^{2}}\frac{1}{r_{1}^{2}}\left( \frac{\alpha_{s}(r_{1}^{2})}{\alpha_{s}(r_{2}^{2})}- 1\right), \\
& K_{\rm rc}^3 = \frac{N_{c}\alpha_{s}(r^{2})}{2\pi^{2}}\frac{1}{r_{2}^{2}}\left( \frac{\alpha_{s}(r_{2}^{2})}{\alpha_{s}(r_{1}^{2})}- 1\right), 
\end{align}
whereas the running coupling kernel is then given by the addition of these constituent terms as
\begin{equation}
    K_{\rm rc} = K_{\rm rc}^1 + K_{\rm rc}^2 + K_{\rm rc}^3.
\end{equation}
The contribution of the three terms is shown in Fig.~\ref{kernel_elements} at $r$ = 1\,GeV$^{-1}$ and $\theta_{rr_1}=\pi/2$ for each of the two kernels.
The fact that the three terms are added in $K_{\rm rc}$, but multiplied in $K_{\rm ci}$ explains numerically the suppression. Even though, the first term is essentially the same for both kernels, the additive character of $K_{\rm rc}$ makes it deviate from the collinearly improved kernel at large $r_1$ values as shown in Fig.~\ref{kernel_elements}. 
There, we can see that even though the kernels are comparable in the low-$r_1$ region, at large  $r_1$ values, the $K_{\rm rc}^2$ and $K_{\rm rc}^3$ terms become dominant, whereas in the collinearly improved kernel, the $K_{\rm ci}^1$ term suppresses the total value.

The physical reason of this suppression can be traced back to the fact that large daughter dipoles do not follow the time-ordering prescription (that is, they would live longer than the parent dipole)  built in when setting up the resummation that leads to the collinearly improved kernel~\cite{Beuf:2014uia,Iancu:2015joa}.

\begin{figure}[!ht]
  \centering
  \includegraphics[width=0.6\linewidth]{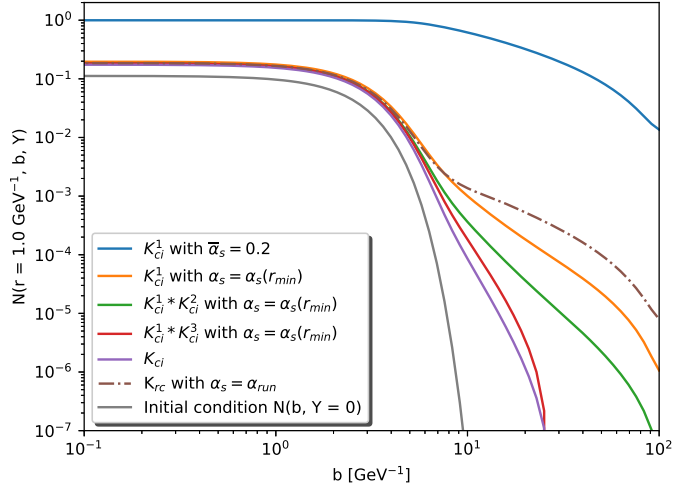}
\caption{The scattering amplitude evolved to $Y = 3$ with various kernels illustrates the effect of the different terms in the evolution and demonstrate that the computation based on the   $K_{\rm ci}$ kernel does not develop   the Coulomb tails seen  when the $K_{\rm rc}$ kernel is used.}
\label{kernel_evolution}
\end{figure}

\subsection{Contribution of the kernel terms to the evolution}
The suppression for large sizes of the daughter dipole in the kernel is translated as a suppression of the amplitude at large $b$ in the evolution. In this region only  large  $r_{1,2}$  contribute to the total integral in Eq.~(\ref{BKused}). This is true because a large impact parameter means that the probing dipole is far away from the target proton and the amplitude is therefore (at the initial condition) exponentially suppressed. Only dipoles with $r_1$ ($r_2$) $\sim$ $2b$ can be oriented so that their impact parameters $b_1$ ($b_2$) are small, such that they contribute to the evolution. But, since $K_{\rm ci}$ is suppressed in this region, the integral will be suppressed as well and the scattering amplitude will not grow fast at large $b$.

This can be numerically studied by computing the contribution to the evolution of the three terms in the collinearly improved kernel. Figure~\ref{kernel_evolution} shows the scattering amplitude  after evolution to $Y=3$ using each time a kernel formed with different constituents. It is clearly seen that the impact parameter profile is mostly influenced by the inclusion of the $K_{\rm ci}^{3}$ term with the Bessel functions. This term originates from  resumming double collinear logarithms. Note that  also the term $K_{\rm ci}^{2}$, resumming single collinear logarithms, suppresses the large $b$ region.

\subsection{Behaviour of the solution to BK equation}
\label{ciBK}

\begin{figure}[!ht]
  \centering
   \includegraphics[width=0.49\linewidth]{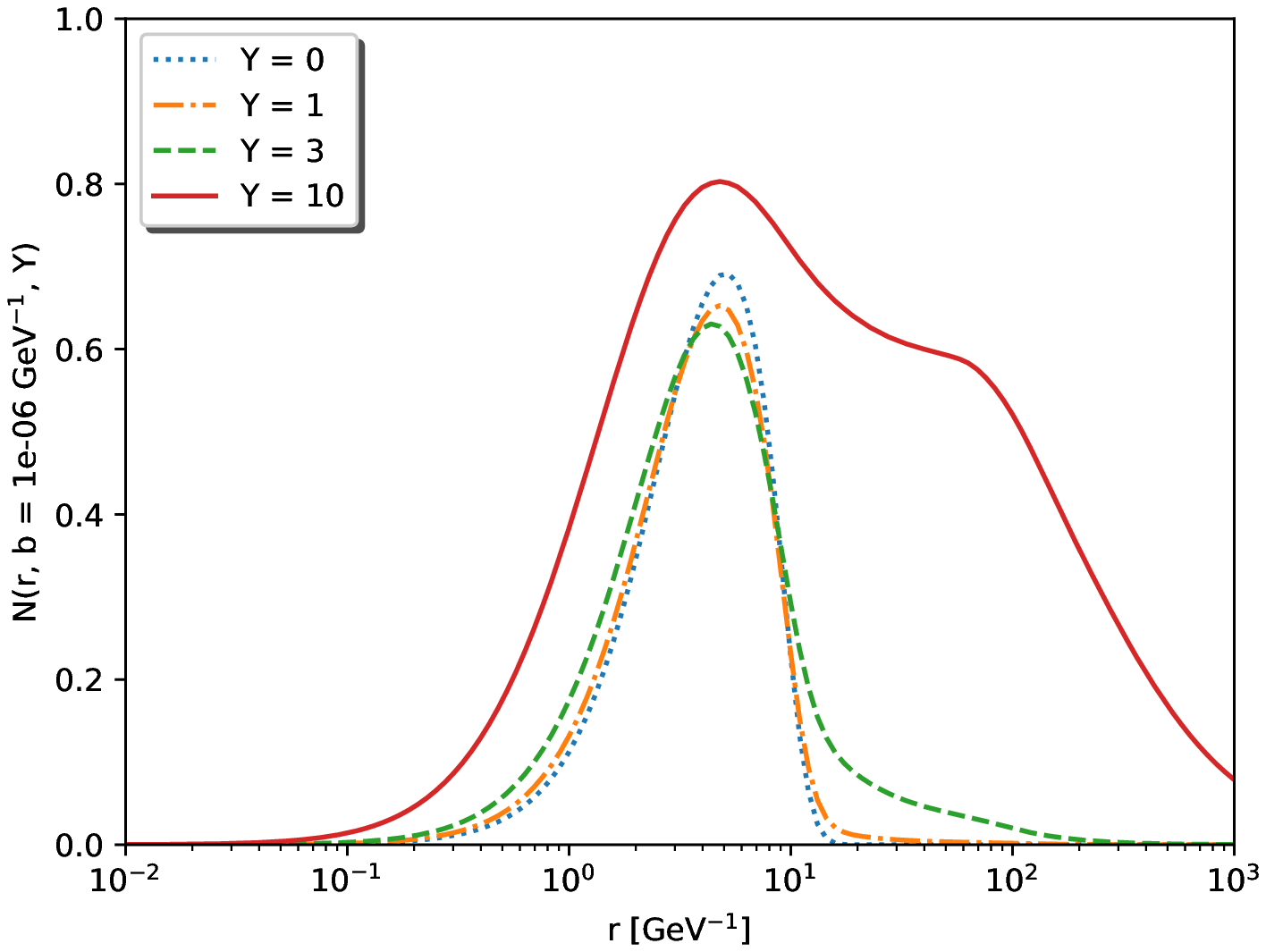}
  \includegraphics[width=0.49\linewidth]{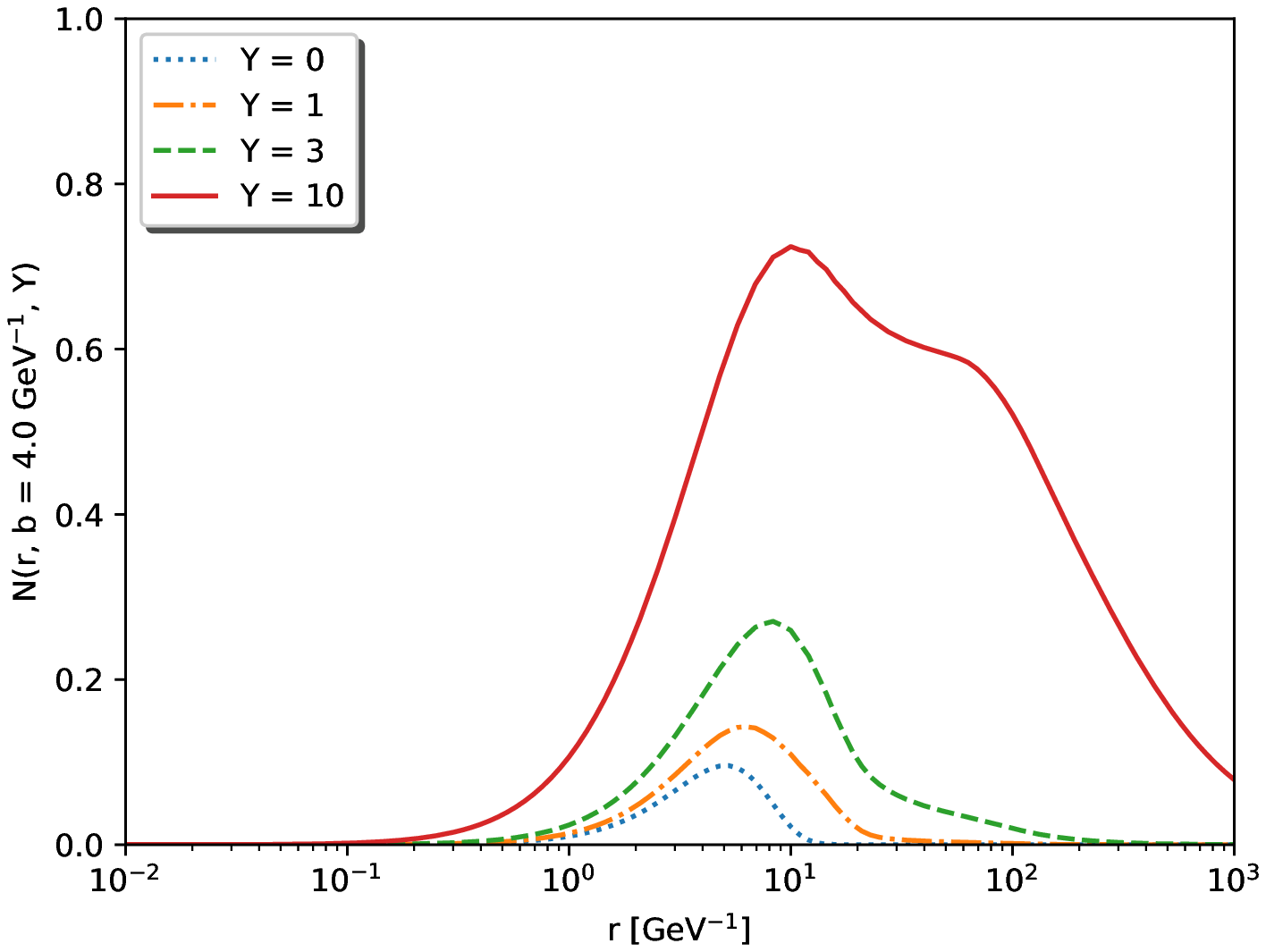}
    \includegraphics[width=0.49\linewidth]{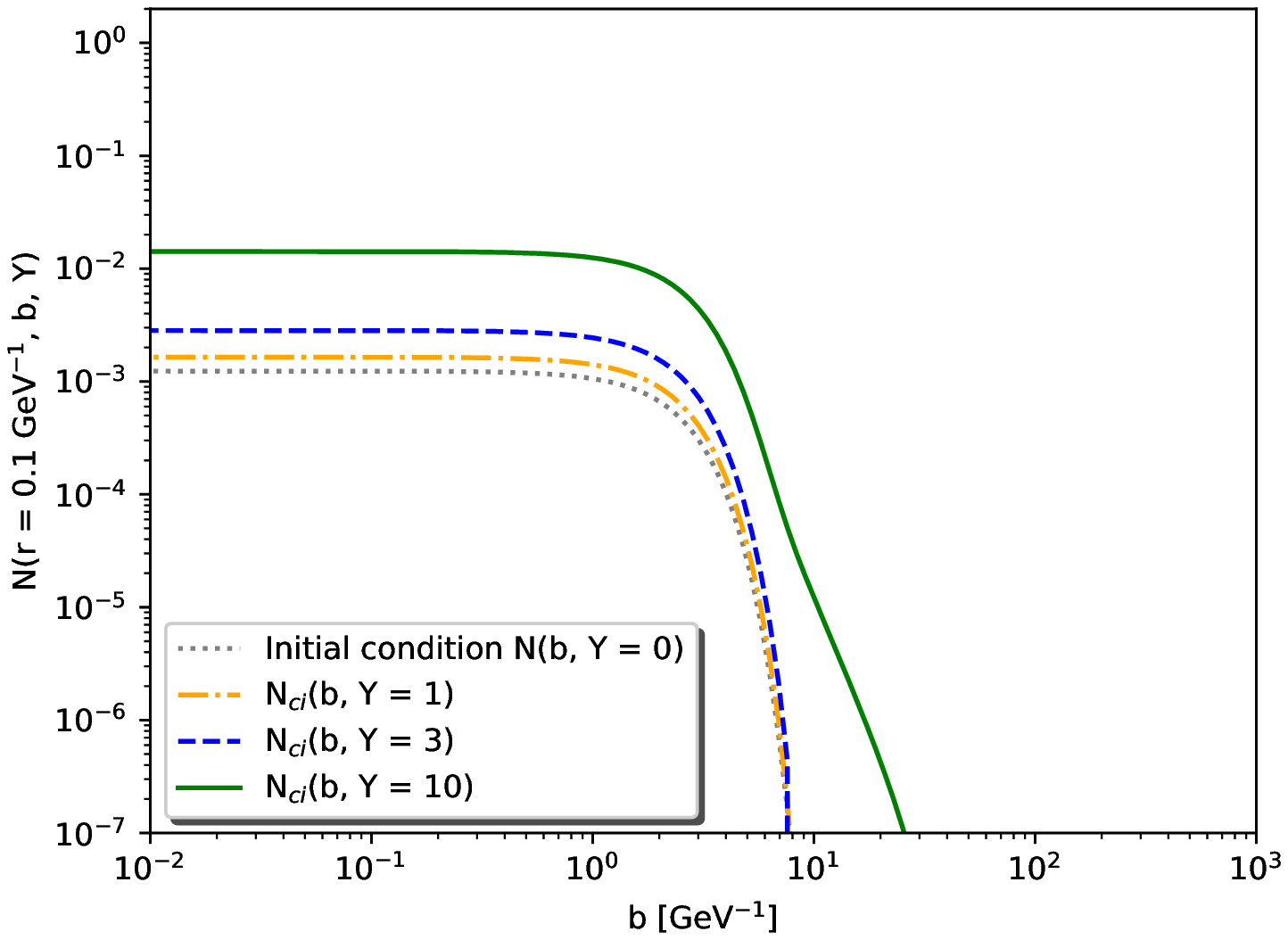}
  \includegraphics[width=0.49\linewidth]{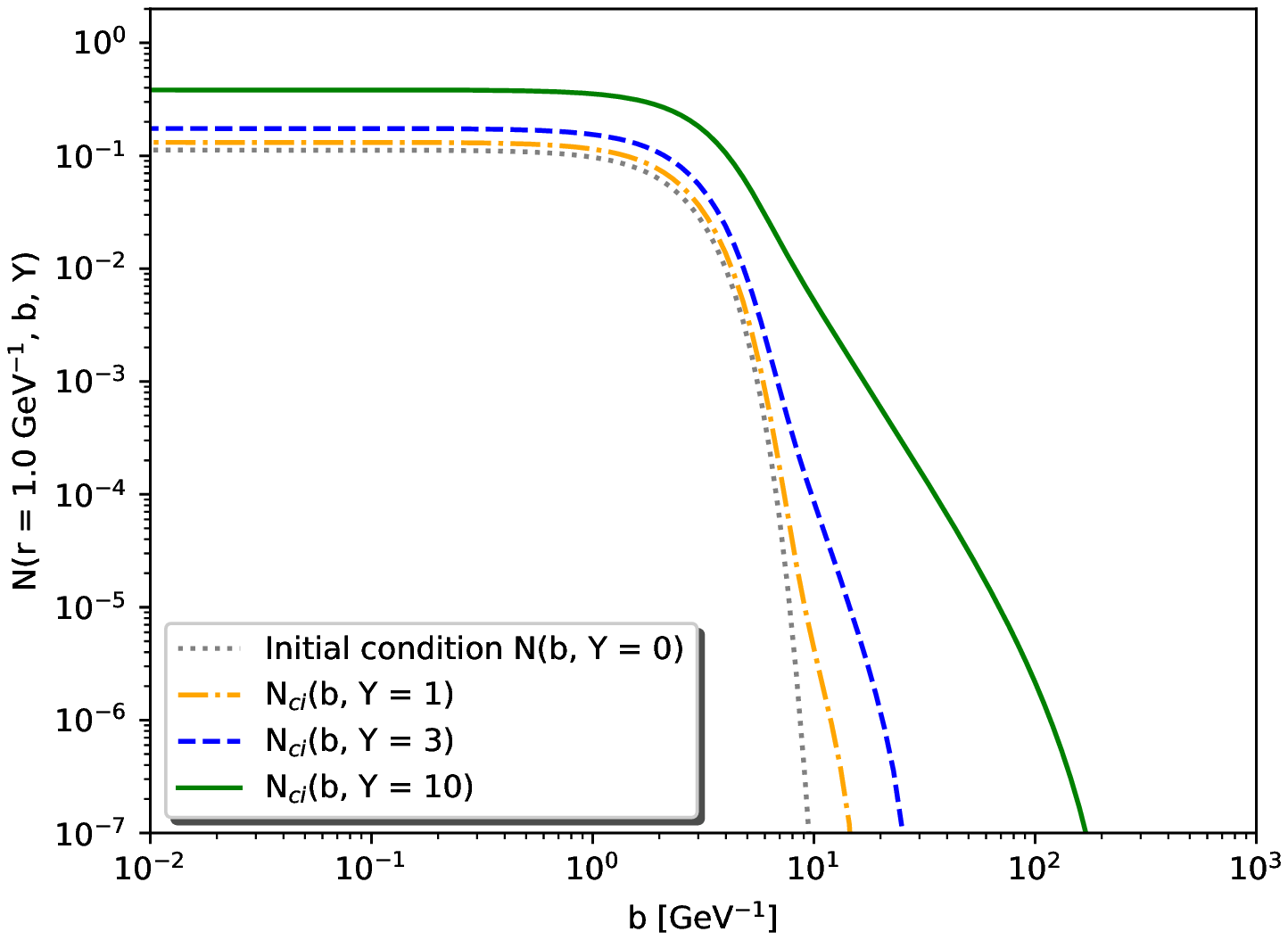}
\caption{The scattering amplitude as a solution to the BK equation with the collinearly improved kernel as a function of $r$ for $b$ = 10$^{-6}$\,GeV$^{-1}$ (upper left) and $b$ = 4\,GeV$^{-1}$ (upper right), and as a function of $b$ at $r$ = 0.1\,GeV$^{-1}$ (lower left) and at $r$ = 1\,GeV$^{-1}$ (lower right).}
\label{bdepb_vs_r}
\end{figure}

\begin{figure}[!ht]
  \centering
  \includegraphics[width=0.49\linewidth]{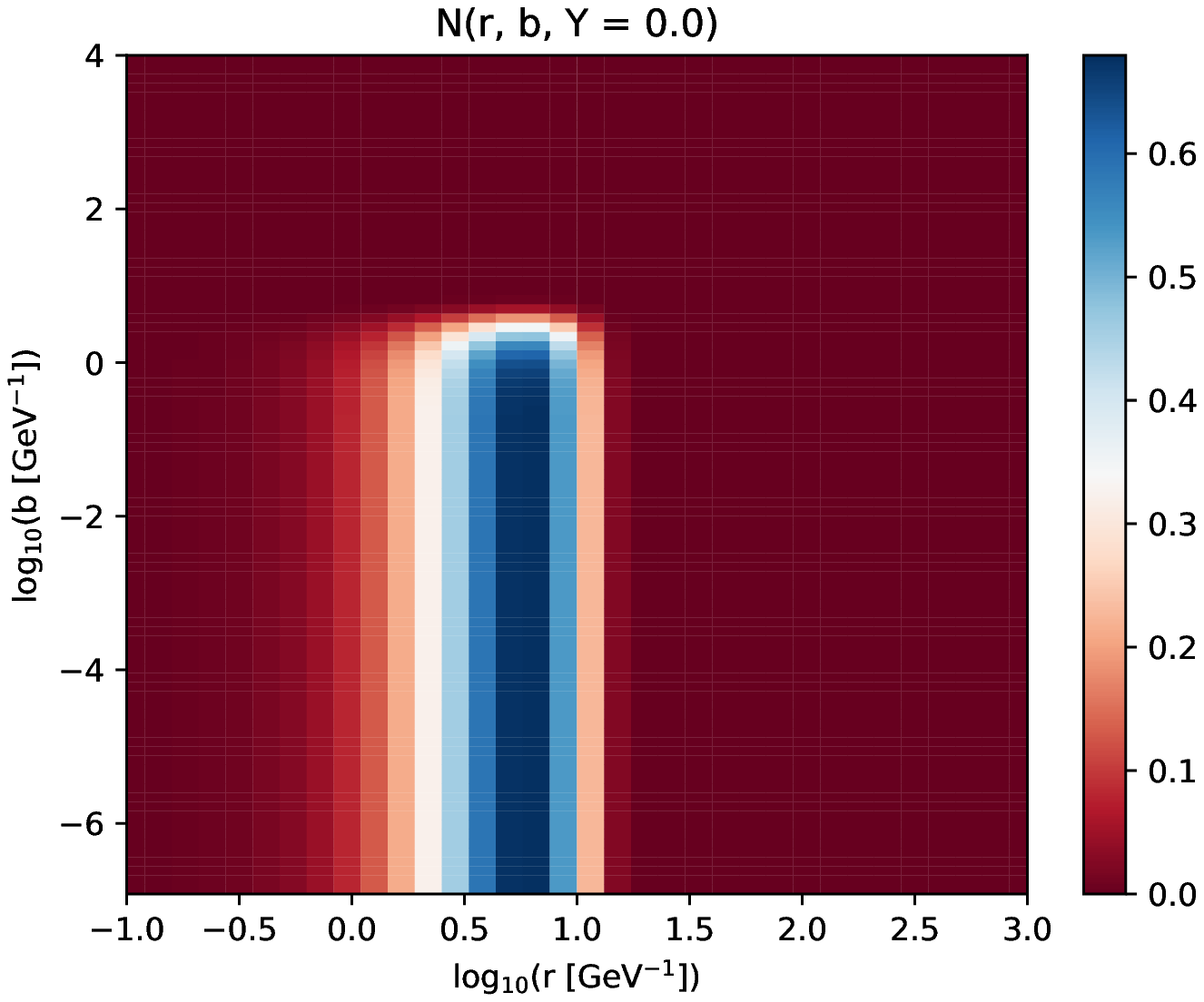}
  \includegraphics[width=0.49\linewidth]{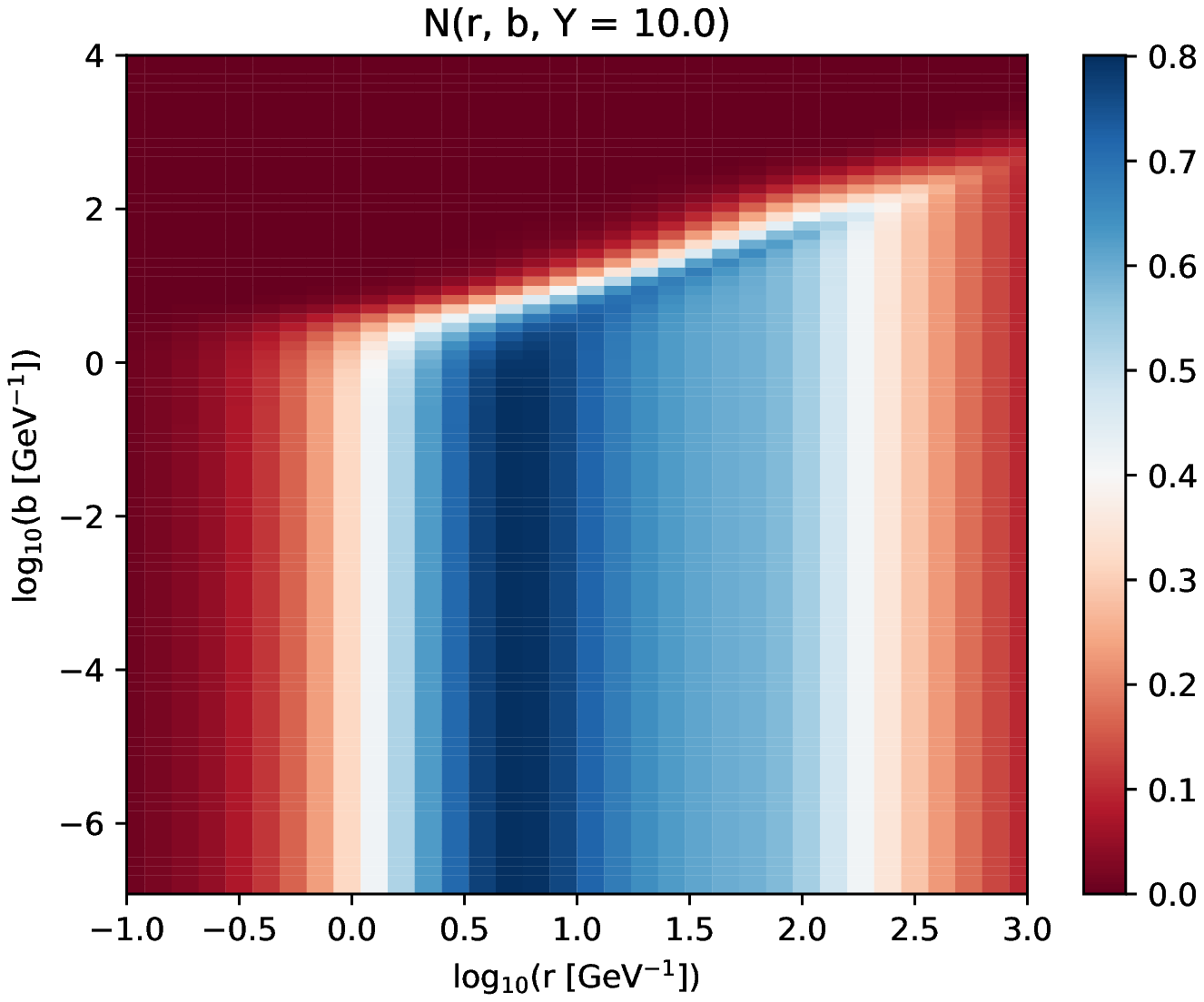}
\caption{Evolution of the scattering amplitude  from the initial condition at  $Y$ = 0 (left) to $Y$ = 10 (right).}
\label{bdepb_2d}
\end{figure}

The evolution of the scattering amplitude as a function of $r$ for different fixed values of $b$ is shown in the upper panels of Fig.~\ref{bdepb_vs_r}, while the lower panels of the same figure show the evolution as a function of $b$ for two fixed values of $r$. A two-dimensional view of the amplitude at two stages of the evolution is shown in Fig.~\ref{bdepb_2d}.
 The amplitude decreases fast for  small dipole sizes as expected. The suppression of large  dipole sizes imposed in the initial condition is lifted with evolution. Similar behaviour was observed in previous studies, e.g.~\cite{Berger:2010sh}. Nonetheless, in the case of the collinearly improved kernel the growth at the largest dipole sizes is not as fast and a shoulder appears, after which the amplitude is again suppressed. The behaviour as a function of impact parameter has been discussed above; the profile impact parameter grows, but the development of Coulomb tails is suppressed. Recently, a similar finding has been reported for the case of NLO BFKL equations at large impact parameters~\cite{Contreras:2019vox}.  

\begin{figure}[!ht]
  \centering
  \includegraphics[width=0.6\linewidth]{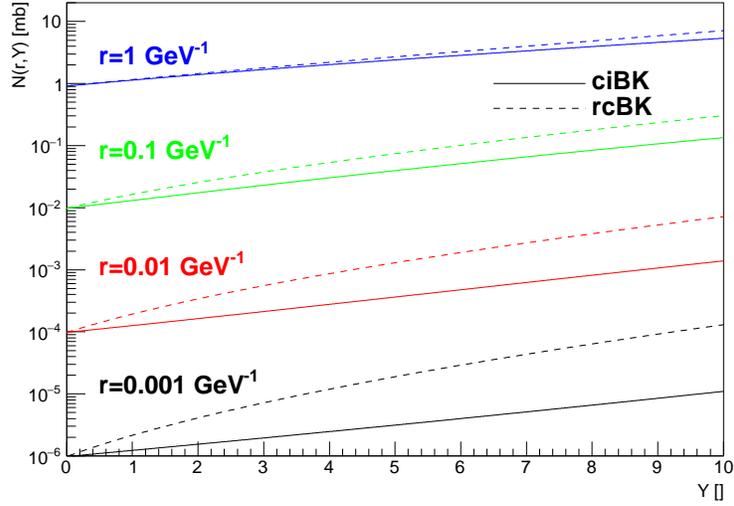}
\caption{Growth of the  dipole-target amplitude integrated over impact parameter as a function of rapidity for solutions of the BK equations with the running coupling and the collinearly improved kernel.}
\label{grow}
\end{figure}

Finally, Fig.~\ref{grow} shows $N(r,Y)$, defined as
\begin{equation}
N(r,Y) = \int d^2\vec{b} N(r,b,Y),
\end{equation}
for different dipole sizes and for two kernels, the running coupling and the collinearly improved. 
For small dipoles the difference is larger and it grows with rapidity. At larger dipole sizes the difference between both kernels is smaller. Note that for the comparisons to data discussed below, the main numerical contribution comes from the region of relatively large dipoles. For the case of the structure function the main contribution for virtualities between 1 and 10 GeV$^2$ comes from dipoles of sizes on the range around 0.1/GeV to 10/GeV, see e.g. the lower panel of Fig. 4 in~\cite{Cepila:2015qea}.

Another interesting observation is that  $N(r,Y)$ is related to the $\sigma_0$ parameter introduced in studies based on the rcBK equation without impact-parameter dependence. Basically, $\sigma_0$ corresponds to the scale of $N(r,Y)$.  Standard values found for this parameter are a few tens of mb, see e.g. Table 1 in~\cite{Albacete:2010sy}. Figure~\ref{grow} justifies the order of magnitude of these values from the perspective of an impact-parameter dependent computation. 

\section{Deep Inelastic Scattering}
\label{Data}

\subsection{Structure function and reduced cross section}
\label{StructureFunction}

Due to the fact, that the dipole lives much longer than the typical interaction time, the computation of the  total deep-inelastic scattering (DIS) cross section can be written as the convolution of separate terms. One of them is the wave function representing the probability of a virtual photon splitting into a quark-antiquark dipole. Here formulas and notation of~\cite{GolecBiernat:1998js} are used:
\begin{equation}
  \mid\Psi_{T}^{i}(z, \vec{r}, Q^{2})\mid^{2} = \frac{3\alpha_{\rm em}}{2\pi^{2}} e_{q_{i}}^{2}\Big((z^{2} + (1-z)^{2})\epsilon^{2}K^{2}_{1}(\epsilon r) + m_{q_{i}}^{2}K^{2}_{0}(\epsilon r)\Big),
\end{equation}
 and 
\begin{equation}
  \mid\Psi_{L}^{i}(z, \vec{r}, Q^{2})\mid^{2} = \frac{3\alpha_{\rm em}}{2\pi^{2}}e_{q_{i}}^{2}\Big(4Q^{2}z^{2}(1-z)^{2}K^{2}_{0}(\epsilon r)\Big)
\end{equation}
for the transverse and longitudinal polarisation of the incoming photon, respectively, where $z$ is the fraction of the total longitudinal momentum of the photon carried by the quark, $K_{0}$ and $K_{1}$ are the MacDonald functions, $Q^2$ is the virtuality of the probing photon, $e_{qi}$ is the fractional charge (in units of elementary charge) of quark $i$, $\alpha_{\rm em}$ = 1/137 and
\begin{equation}
 \epsilon^{2} = z(1-z)Q^{2} + m_{q_{i}}^{2},
 \label{epsilon}
\end{equation}
where $m_{q_{i}}$ is the mass of the considered quark, which is set to 100\,MeV$/c^2$ for light quarks and 1.3\,GeV$/c^2$ for charm quark and 4.5\,GeV$/c^2$ for bottom quark. Note that the computed structure function does not depend strongly on the value of the mass of the light quarks (as was reported in~\cite{Iancu:2015joa}); this has been checked by also using $m_{u,d,s}$ = 10\,MeV$/c^2$, which did not influence the description of data. The total wave function then is

\begin{equation}\label{psi}
 \mid\Psi_{T,L}^{i}( z, \vec{r}\,)\mid^{2} =\mid \Psi^{i}_{T}(z, \vec{r}\,)\mid ^{2} +  \mid\Psi^{i}_{L}(z, \vec{r}\,)\mid^{2}.
\end{equation}

According to the optical theorem, one can link the dipole--target cross section to the scattering amplitude by
\begin{equation}
\frac{\mathrm{d}\sigma^{q\bar{q}}(\vec{r},x)}{\mathrm{d} \vec{b}} = 2N(\vec{r}, \vec{b}, x).
\label{dipole-cs}
\end{equation}

Furthermore, it is usual to shift the value of the $x$ at which the structure function and reduced cross section are computed according to the photoproduction kinematic shift~\cite{GolecBiernat:1998js}
\begin{equation}
\tilde{x} = x \left( 1 + \frac{4m^2_{q_i}}{Q^2} \right).
\end{equation}

Using these ingredients, the relation for the computation of the structure function in the dipole model framework is
\begin{equation}\label{F22}
 F_{2}(x, Q^{2}) = \frac{Q^{2}}{4\pi^{2}\alpha_{\rm em}}\int\sum_{i} d\vec{r}d\vec{b}dz \mid\Psi_{T,L}^{i}( z, \vec{r}\,)\mid^{2} \frac{\mathrm{d}\sigma^{q\bar{q}}(\vec{r},\tilde{x})}{\mathrm{d} \vec{b}},
\end{equation}
and the reduced cross section is computed as 
\begin{equation}\label{sigma}
 \sigma_{\rm red}(y, x, Q^{2}) = F_{2}(x, Q^{2}) - \frac{y^2}{1 + (1-y)^2}F_{L}(x, Q^{2}),
\end{equation}
where $y=Q^2/(sx)$ is the inelasticity of the process, $s$ is the squared of the centre-of-mass energy of the collision and $F_{L}(x, Q^{2})$ is given by the relation
\begin{equation}\label{FL}
 F_{L}(x, Q^{2}) = \frac{Q^{2}}{4\pi^{2}\alpha_{\rm em}}\int\sum_{i} d\vec{r}d\vec{b}dz \mid\Psi_{L}^{i}( z, \vec{r}\,)\mid^{2} \frac{\mathrm{d}\sigma^{q\bar{q}}(\vec{r},\tilde{x})}{\mathrm{d} \vec{b}}.
\end{equation}

\subsection{Comparison to HERA data}
\label{Dataplots}

\begin{figure}[!ht]
  \centering
   \includegraphics[width=0.65\linewidth]{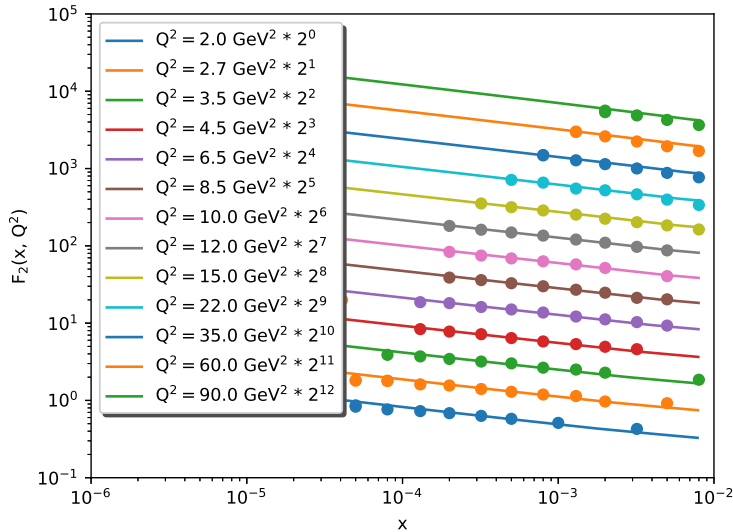}
\caption{Comparison of the structure function data from HERA~\cite{Aaron:2009aa} (solid circles) to the prediction of the impact-parameter dependent BK equation with the collinearly improved kernel (lines).}
\label{bdepbF2}
\end{figure}

\begin{figure}[!ht]
  \centering
   \includegraphics[width=0.49\linewidth]{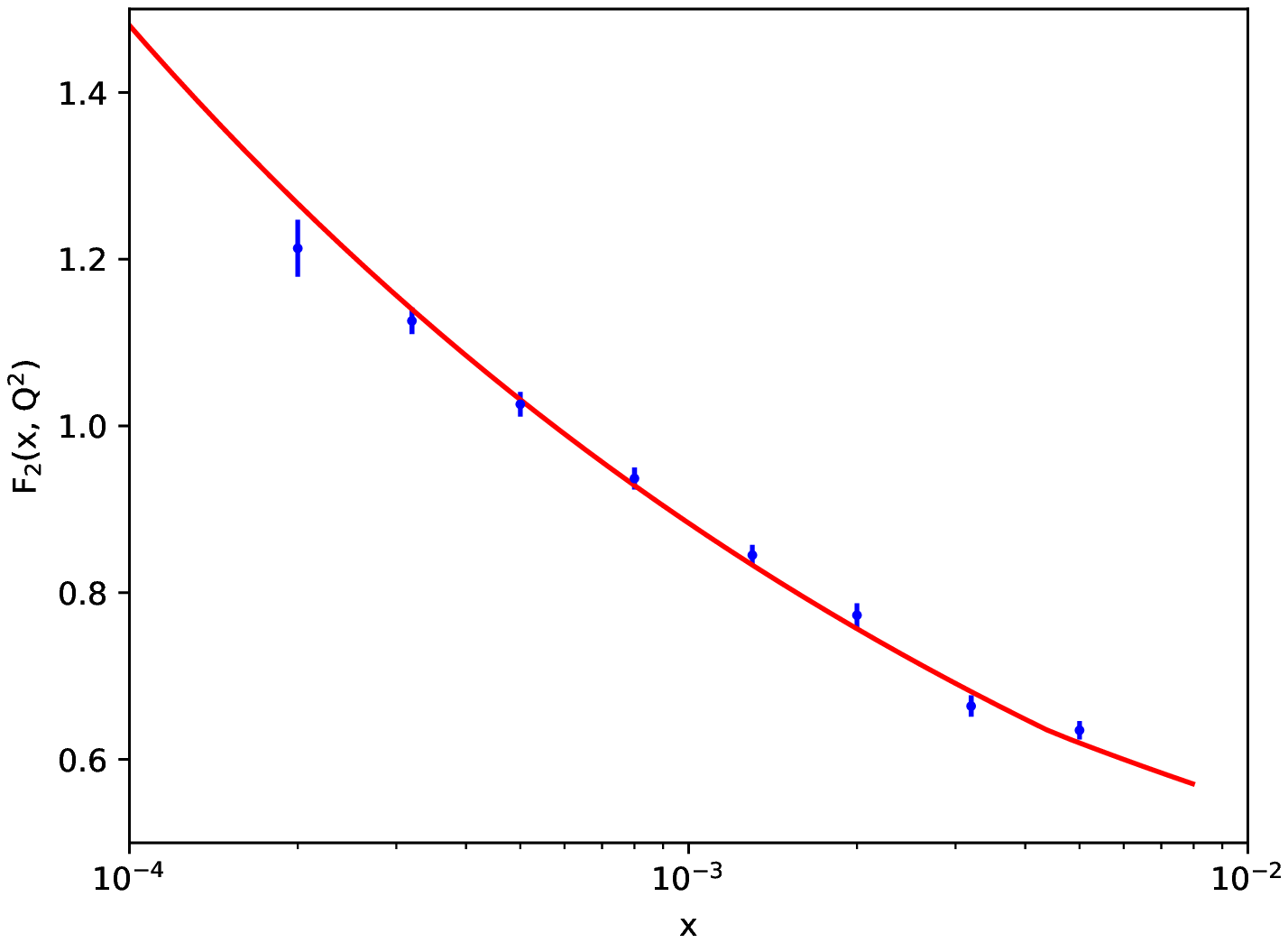}
   \includegraphics[width=0.49\linewidth]{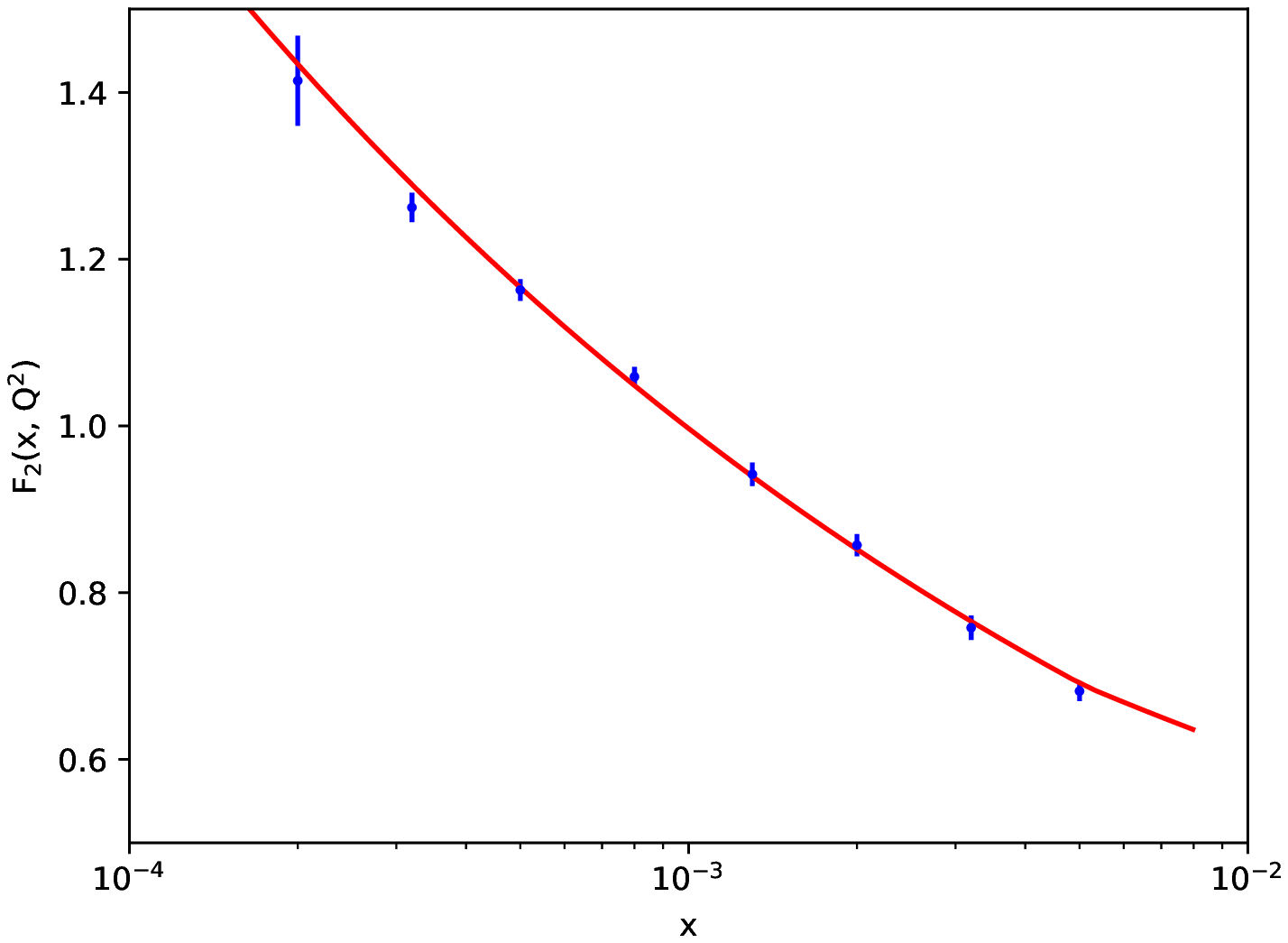}
\caption{Close-up comparison of the structure function data from HERA~\cite{Aaron:2009aa} (blue points) to the b-dependent prediction (red line) for $Q^2$ = 8.5\,GeV$^2$ (left) and $Q^2$ = 12\,GeV$^2$ (right).}
\label{F2_detail}
\end{figure}

\begin{figure}[!ht]
  \centering
   \includegraphics[width=\linewidth]{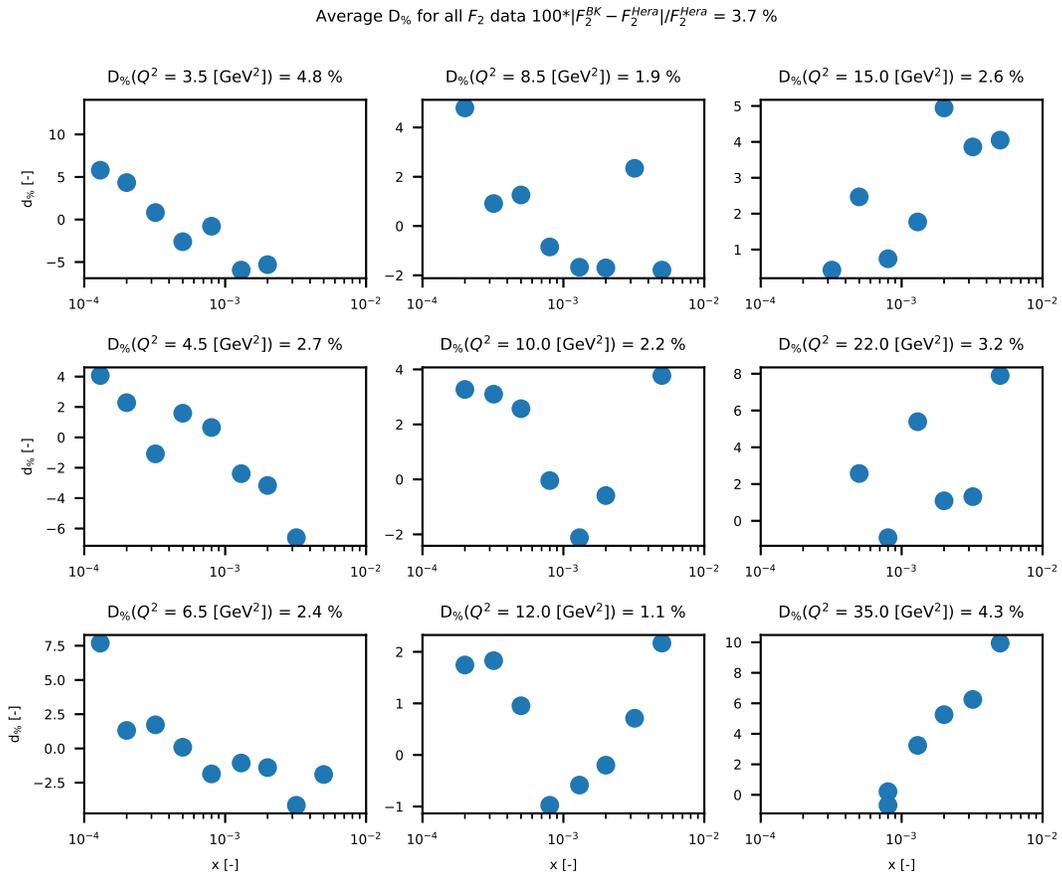}
\caption{The percentage pulls for various values of $Q^2$ and their average value.}
\label{pulls}
\end{figure}

\begin{figure}[!ht]
  \centering
   \includegraphics[width=0.65\linewidth]{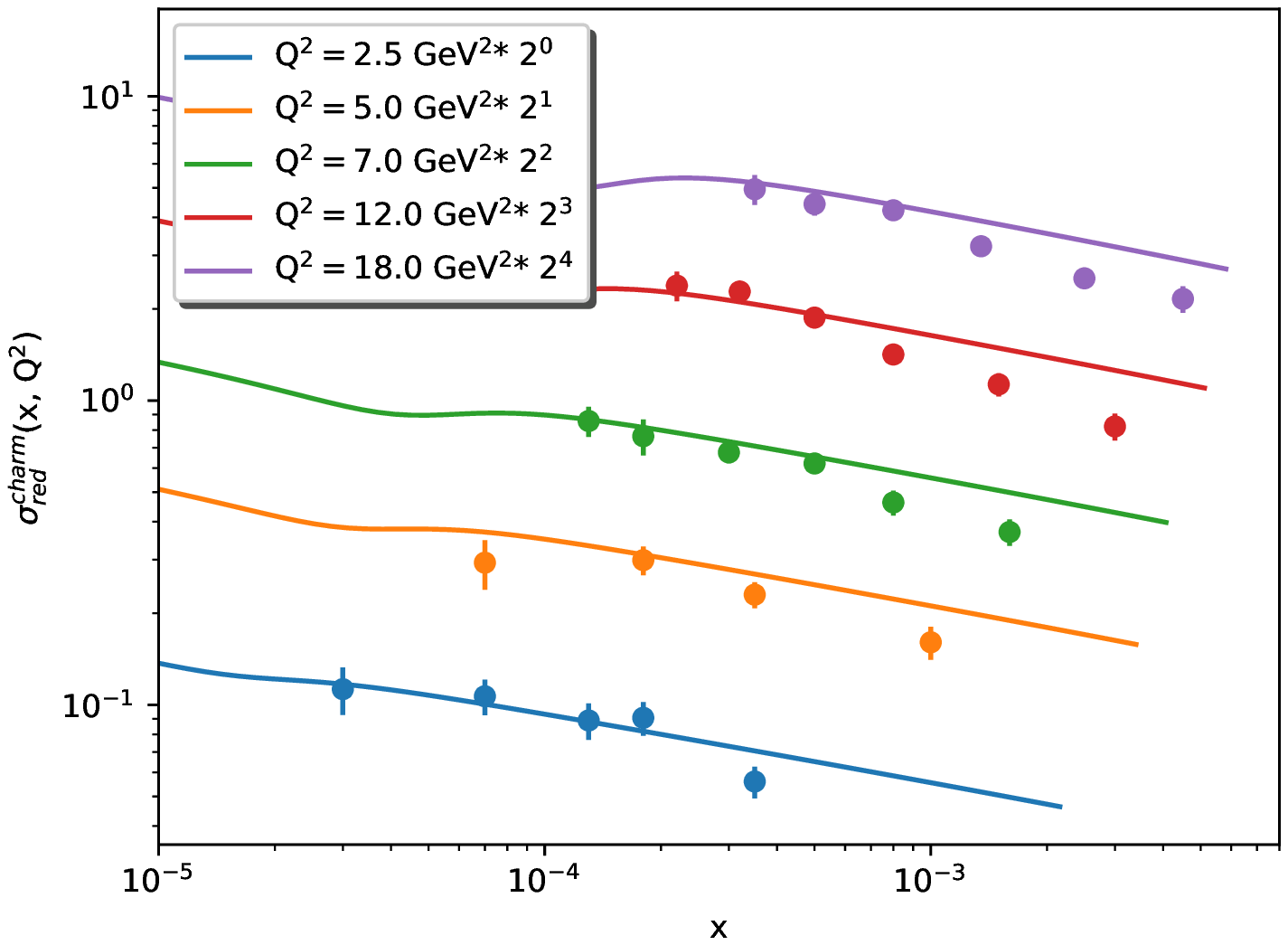}
\caption{The comparison of the prediction for the reduced cross section for  charm to data from HERA~\cite{Aaron:2009aa}.}
\label{bsigma}
\end{figure}

The predictive power of this model is evaluated by  confronting it with data from HERA  on the $F_2(x,Q^2)$ structure function~\cite{Aaron:2009aa} in Fig.~\ref{bdepbF2}. A closer look is given in Fig.~\ref{F2_detail} for two values of the photon virtuality. 
To quantify the level of agreement between data and model, Fig.~\ref{pulls} presents the percentage pulls associated with the structure function, which are given by
\begin{equation}\label{pull}
    d_{\%} = 100\frac{F_2^{\rm BK}(x, Q^2) - F_2^{\rm HERA}(x, Q^2)}{F_2^{\rm HERA}(x, Q^2)}
\end{equation}
 and by $D_{\%}$, which denotes the average of the corresponding values of $d_{\%}$. Finally, for completeness Fig.~\ref{bsigma} shows the comparison of the model and data for the charm component of the proton structure function measured at HERA~\cite{Aaron:2009aa}.

Overall, the agreement between prediction and data is within a few percent over most of the phase space. For our purposes this level of agreements is satisfactory. First, the equation we are using does not include the full angular dependence. Second, we have not needed to add any ad-hoc component to describe data and the values of the parameters are reasonable from the point of view of the physics that is being probed.  Furthermore, note that the BK equation that we are using is definitely not the last word on the subject. The full equation at NLO has already been computed~\cite{Balitsky:2008zza}, and a large effort is being done to use it for phenomenology~\cite{Lappi:2016fmu,Beuf:2014uia,Beuf:2017bpd,Ducloue:2017ftk,Hanninen:2017ddy}. There are also recent developments regarding the most adequate variable to evolve the scattering amplitude~\cite{Ducloue:2019ezk}.


\section{Production of vector mesons}
\label{VM}
\subsection{Exclusive cross section in the colour-dipole approach}
Similarly to the DIS process described in the previous section, the diffractive production of a vector meson as a result of the interaction of a virtual photon with the proton can be  treated within the 
colour-dipole approach. In this formalism, the exclusive cross section to produce a vector meson $\mathrm{V}$ is given by
\begin{equation}
\frac{\mathrm{d}\sigma^{\gamma^*p \rightarrow {\rm V}p}}{\mathrm{d}|t|} \bigg| _{T,L} = \frac{\left(1+\beta^2\right) \left(R_g ^{T,L}\right)^2}{16\pi} |  \mathcal{A}_{T,L}  |^2,
\label{VM-cs-diff-excl}
\end{equation}
where $\mathcal{A}_{T,L}$ is the scattering amplitude of the process. It is given as a convolution of the overlap of photon-meson wave functions with the dipole cross section given in Eq.~(\ref{dipole-cs}) (for a detailed derivation see e.g.,~\cite{Kowalski:2006hc}) and takes the following form
\begin{equation}
\mathcal{A}_{T,L}(x,Q^2,\vec{\Delta}) = i \int \mathrm{d}\vec{r} \int \limits_0^1 \frac{\mathrm{d}z}{4\pi} \int \mathrm{d}\vec{b} |\Psi_{\rm V}^* \Psi_{\gamma^*}|_{T,L} \exp \left[ -i\left( \vec{b} - (1-z)\vec{r} \right)\vec{\Delta} \right] \frac{\mathrm{d}\sigma^{q\bar{q}}}{\mathrm{d} \vec{b}},
\label{VM-amplitude}
\end{equation}
where the subscripts $T$, $L$ denote the contribution from the virtual photon with transverse, respectively longitudinal, polarisation, $\Psi_{\gamma^*}$ is the wave function of a virtual photon which fluctuates into a dipole, $\Psi_{\mathrm{V}}$ represents the wave function of the vector meson, and $\vec{\Delta}^2 \equiv -t$, the square of the four momentum transferred in the proton vertex. Under the assumption of large photon-proton centre-of-mass energies $\Wgp$, 
\begin{equation}
x = \frac{Q^2 + M^2}{\Wgp^2 + Q^2},
\label{x}
\end{equation}
where $M$ is the mass of the given vector meson.

The wave functions of a vector meson are modelled under the assumption that the vector meson is predominantly  a $q\bar{q}$ pair with the same polarisation and the spin structure as the original photon. The overlap of the photon-meson wave functions in Eq.~(\ref{VM-amplitude}) is given as
\begin{equation}
|\Psi_{\rm V} ^* \Psi_{\gamma^*}|_T = \hat{e}_f e \frac{N_C}{\pi z(1-z)} \left[ m_f ^2 K_0 (\epsilon r) \phi_T (r,z) - \left( z^2 + (1-z)^2 \right) \epsilon K_1 (\epsilon r) \partial_r \phi _T (r,z) \right],
\label{VM-psipsiT}
\end{equation}
and
\begin{equation}
|\Psi_{\rm V} ^* \Psi_{\gamma^*}|_L = \hat{e}_f e \frac{N_C}{\pi}2Qz(1-z) K_0(\epsilon r) \left[ M \phi _L (r,z) + \delta \frac{m_f ^2 - \nabla _r ^2}{Mz(1-z)} \phi _L (r,z) \right],
\label{VM-psipsiL}
\end{equation}
with $\hat{e}_f$ being the effective charge of the given vector meson, $\epsilon$ defined by Eq. (\ref{epsilon}), and the parameter $\delta$ is a switch to include ($\delta = 1$) or exclude ($\delta = 0$) the non-local term in the longitudinal contribution. The scalar part $\phi _{T,L}$ of the wave function is, in general, model dependent. For our studies, we use the boosted Gaussian model~\cite{Nemchik:1994fp, Nemchik:1996cw, Forshaw:2003ki} in which the $\delta$ parameter is fixed to one. The values of the parameters for the wave functions of all vector mesons are fixed according to Table I in~\cite{Bendova:2018bbb}.

The total exclusive cross section to produce a vector meson is given by the sum of the transverse and the longitudinal contributions defined by Eq. (\ref{VM-cs-diff-excl}). Moreover two important corrections have to be applied. The derivation of the formula for the exclusive vector meson cross section is performed under the assumption that the scattering amplitude $\mathcal{A}_{T,L}(x,Q^2,\vec{\Delta})$ is purely imaginary. The real part of the amplitude can be accounted for by the extra term $(1+\beta^2)$ in Eq. (\ref{VM-amplitude}), where $\beta$ is the ratio of real to imaginary parts of the scattering amplitude, for details see~\cite{Kowalski:2006hc}. The other correction takes into account that there are two values of $x$ involved in the interaction of the dipole with the proton and one should therefore use the off-diagonal gluon distribution for vector meson production. This effect can be accounted for by multiplying the scattering amplitude by a factor $R_g ^{T,L}$, called the skewedness correction~\cite{Shuvaev:1999ce}.

\subsection{Comparison to data}

Using the model described in this paper, the cross sections for exclusive photo- and electroproduction of $\phi$, $\jpsi$, $\psip$, and $\Uos$ vector mesons are presented at different virtualities of the exchanged photon and they are compared to available experimental data. The presented results are calculated at the scales which allow perturbative treatment of the specific parts of the model. 

\begin{figure}[ht]
  \centering
   \includegraphics[width=0.49\linewidth]{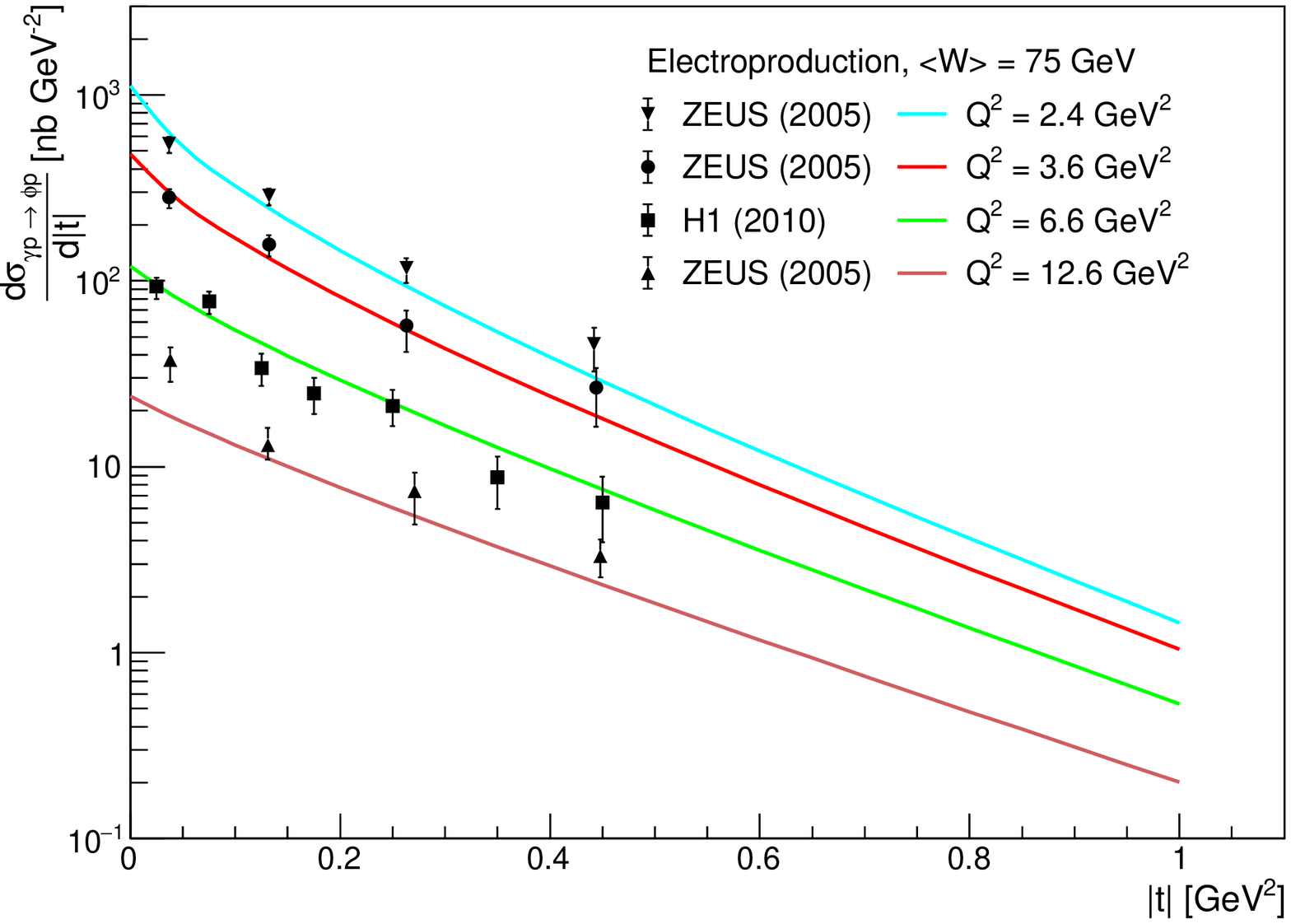}
   \includegraphics[width=0.49\linewidth]{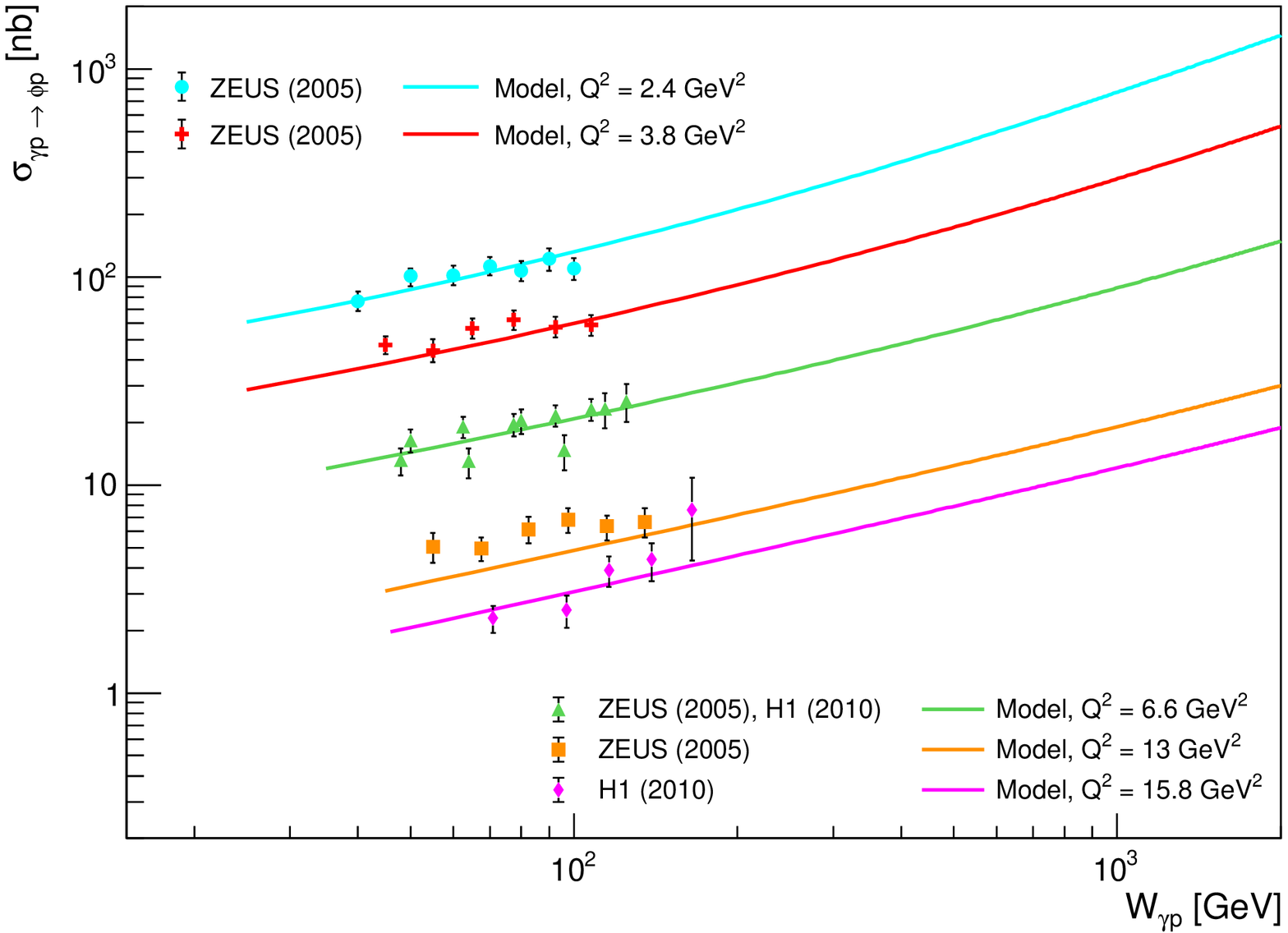}
\caption{Comparison of the predictions of the model (solid lines) with  HERA data from H1~\cite{Aaron:2009xp} and ZEUS~\cite{Chekanov:2005cqa} for the $|t|$ dependence (left) and the $\Wgp$ dependence (right) of the exclusive electroproduction cross section of the $\phi$ meson.}
\label{VM_phi}
\end{figure}

In Fig.~\ref{VM_phi} a comparison of our predictions for the $|t|$-distributions and the total cross sections with HERA H1~\cite{Aaron:2009xp} and ZEUS~\cite{Chekanov:2005cqa} data for the exclusive electroproduction of the $\phi$ meson for several values of $Q^2$ is shown. The predictions give a very good description of the available data, especially at low photon virtualities.

\begin{figure}[ht]
  \centering
   \includegraphics[width=0.49\linewidth]{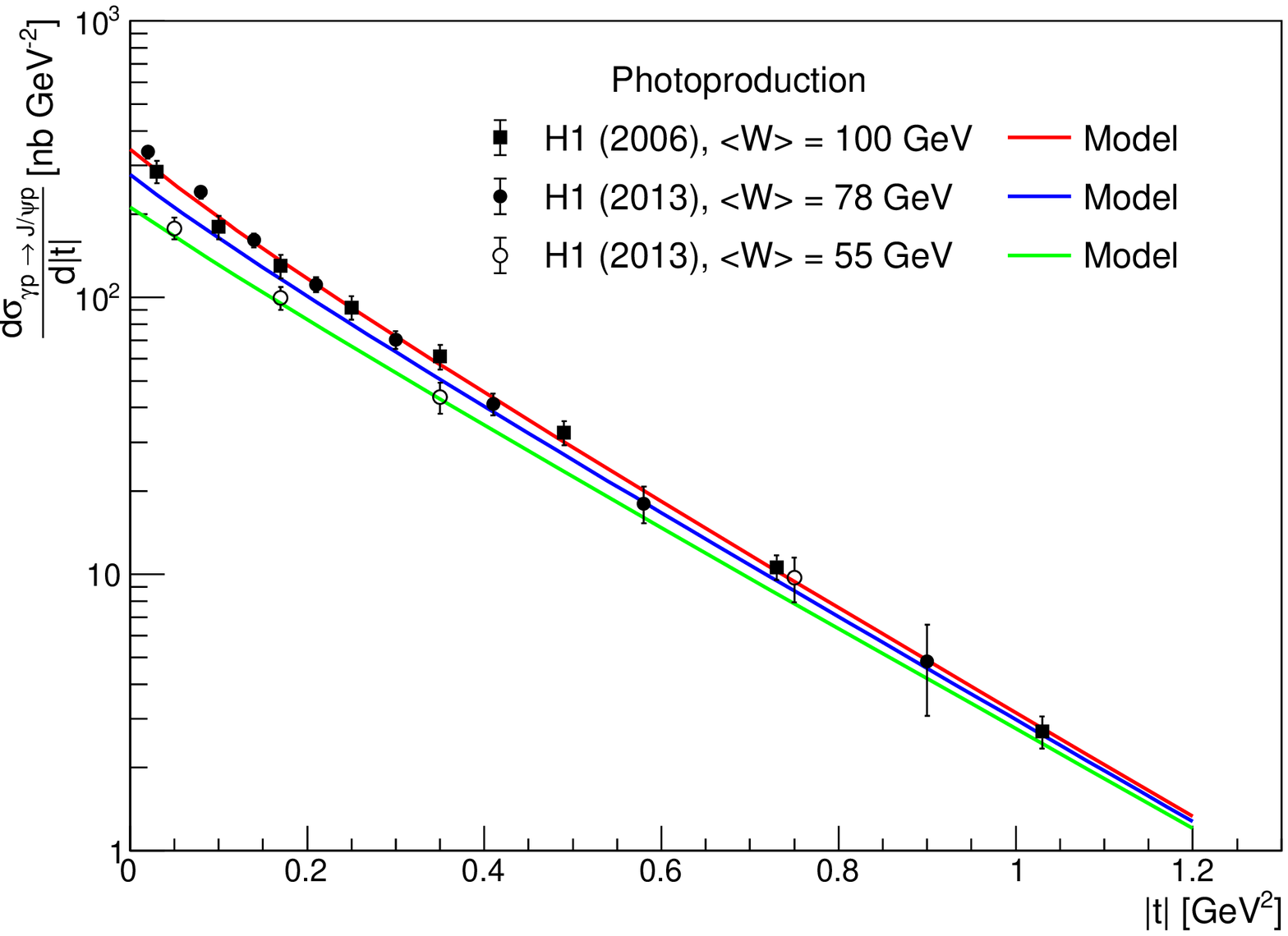}
   \includegraphics[width=0.49\linewidth]{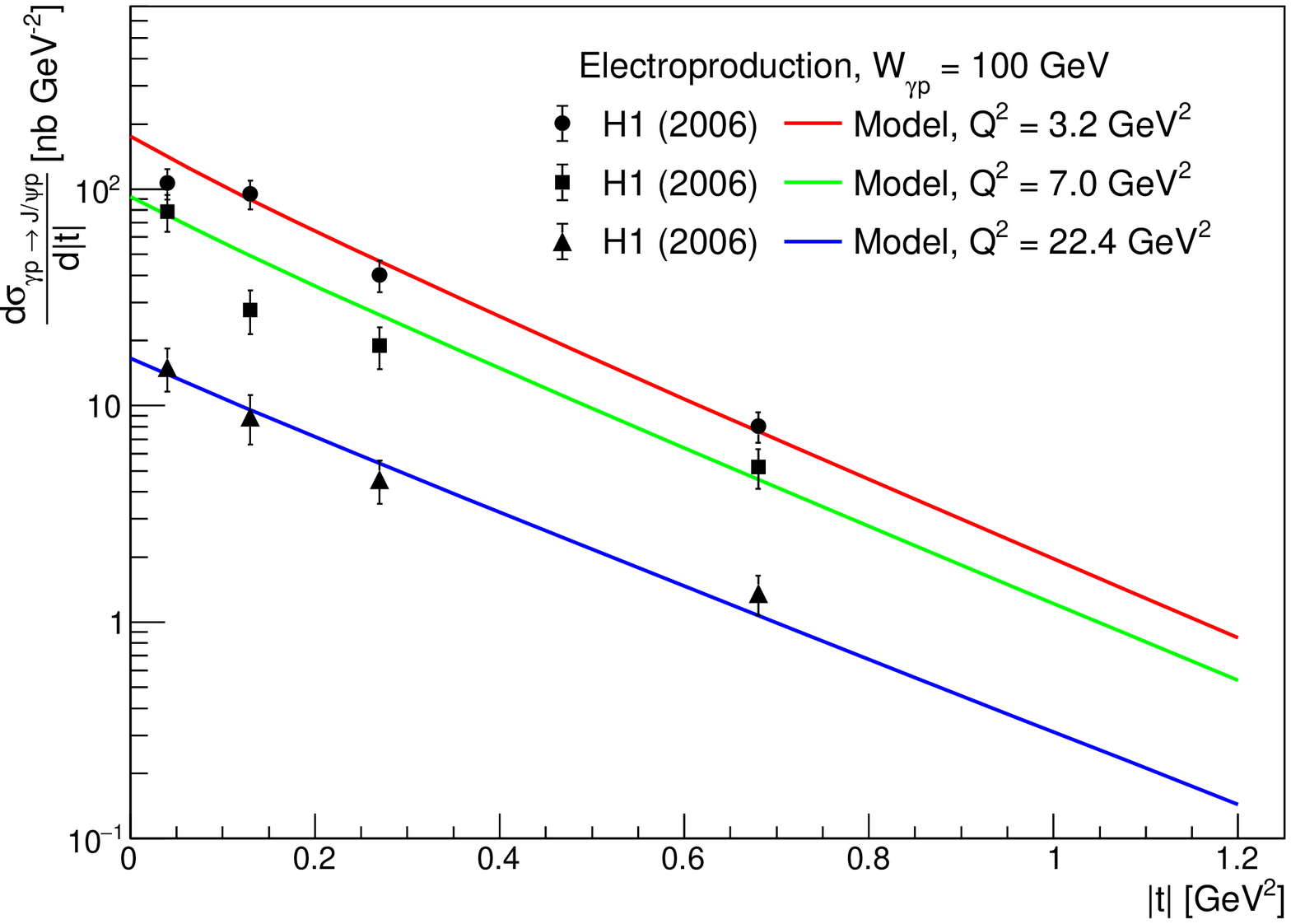}
\caption{Comparison of the predictions of the model (solid lines) with  HERA data from H1~\cite{Aktas:2005xu,Alexa:2013xxa} for the $|t|$ dependence of the exclusive photoproduction (left) and electroproduction (right) cross sections of the $\jpsi$ meson.}
\label{VM_Jpsi_t-distribution}
\end{figure}

\begin{figure}[ht]
  \centering
   \includegraphics[width=0.49\linewidth]{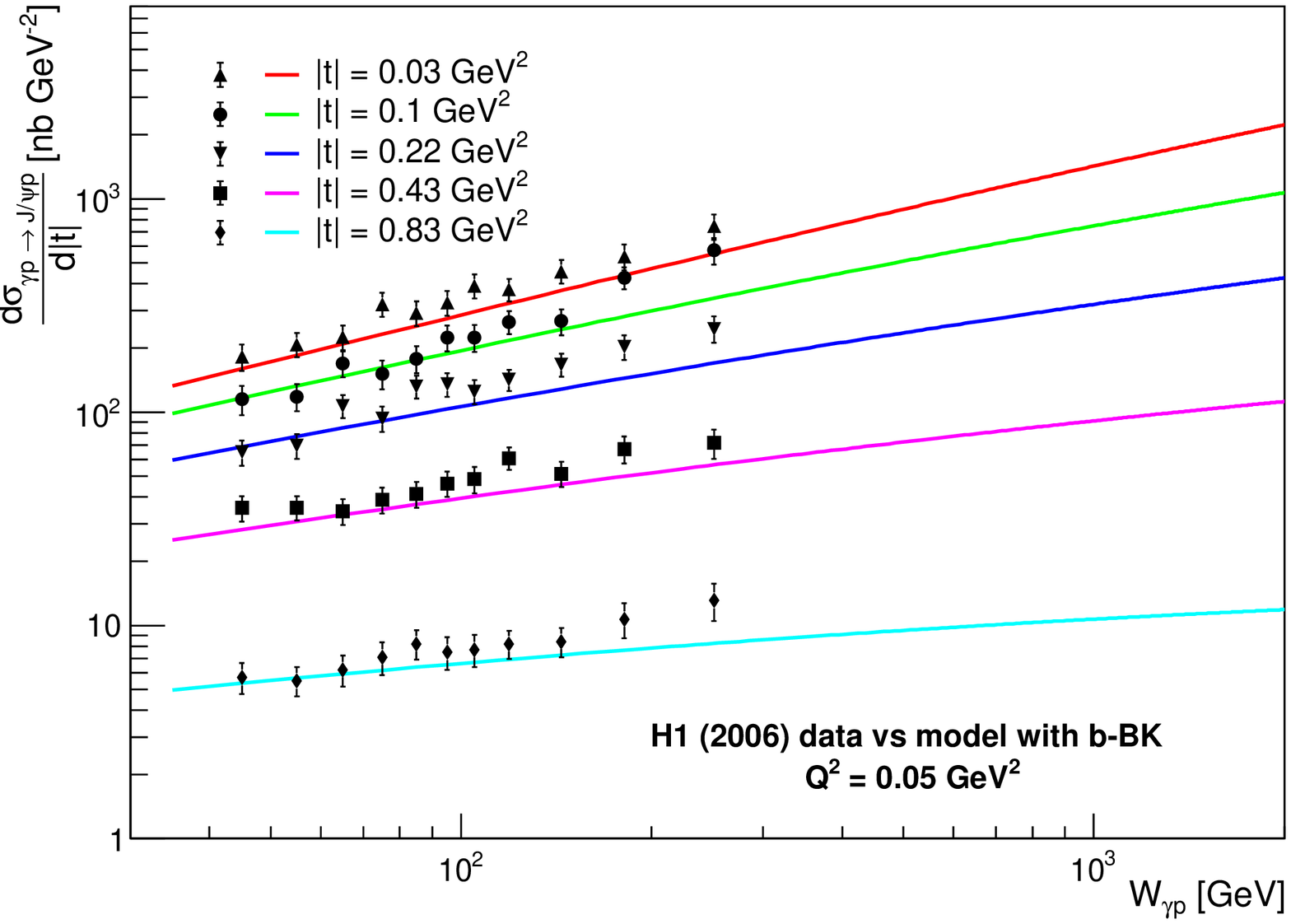}
   \includegraphics[width=0.49\linewidth]{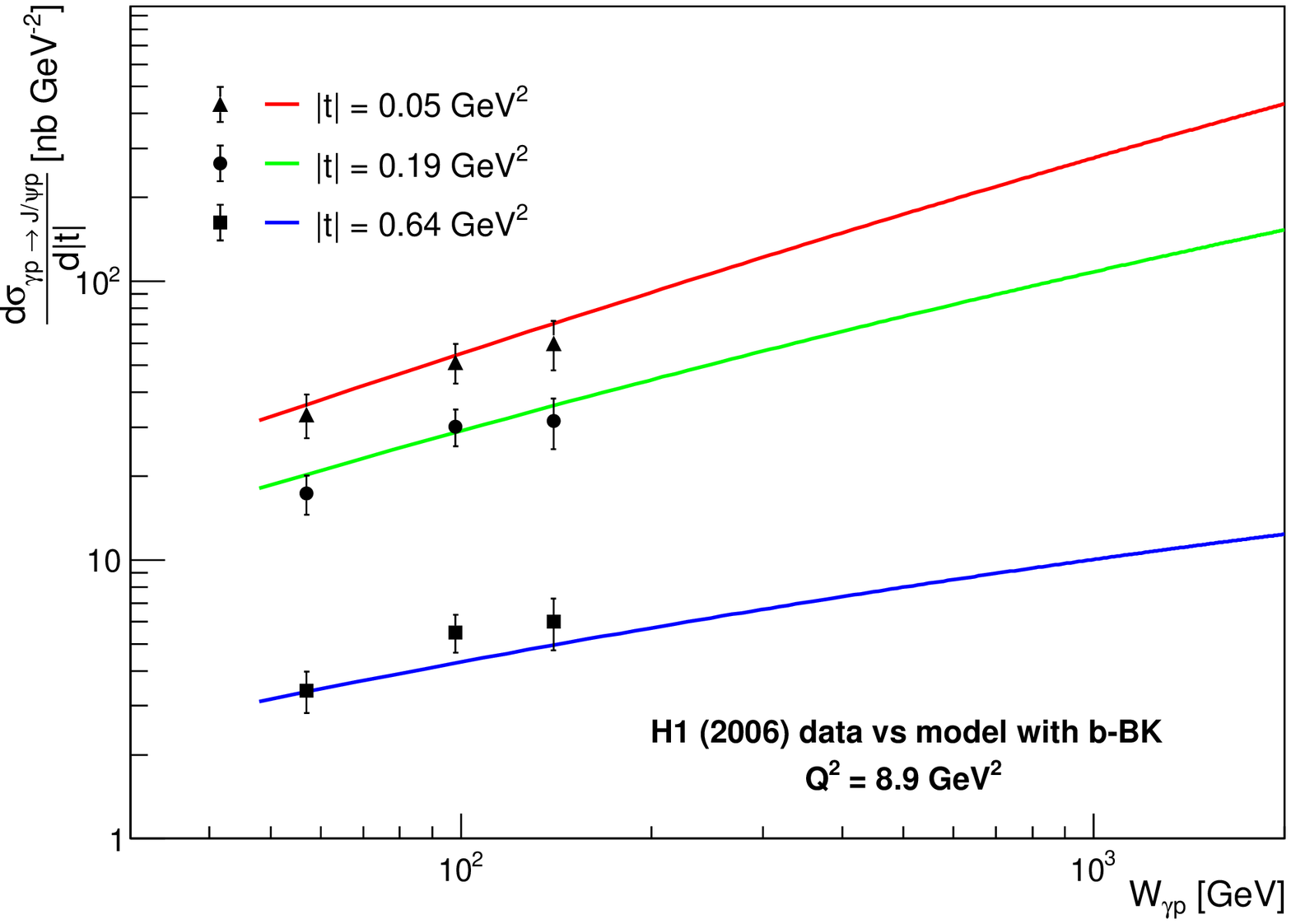}
\caption{Comparison of the predictions of the model (solid lines) with HERA data from H1~\cite{Aktas:2005xu} for the $W$ dependence of the exclusive photoproduction (left) and electroproduction (right) cross sections of the $\jpsi$ meson at fixed $|t|$ values.}
\label{VM_Jpsi_fix-t_W-dependence}
\end{figure}

\begin{figure}[ht]
  \centering
   \includegraphics[width=0.49\linewidth]{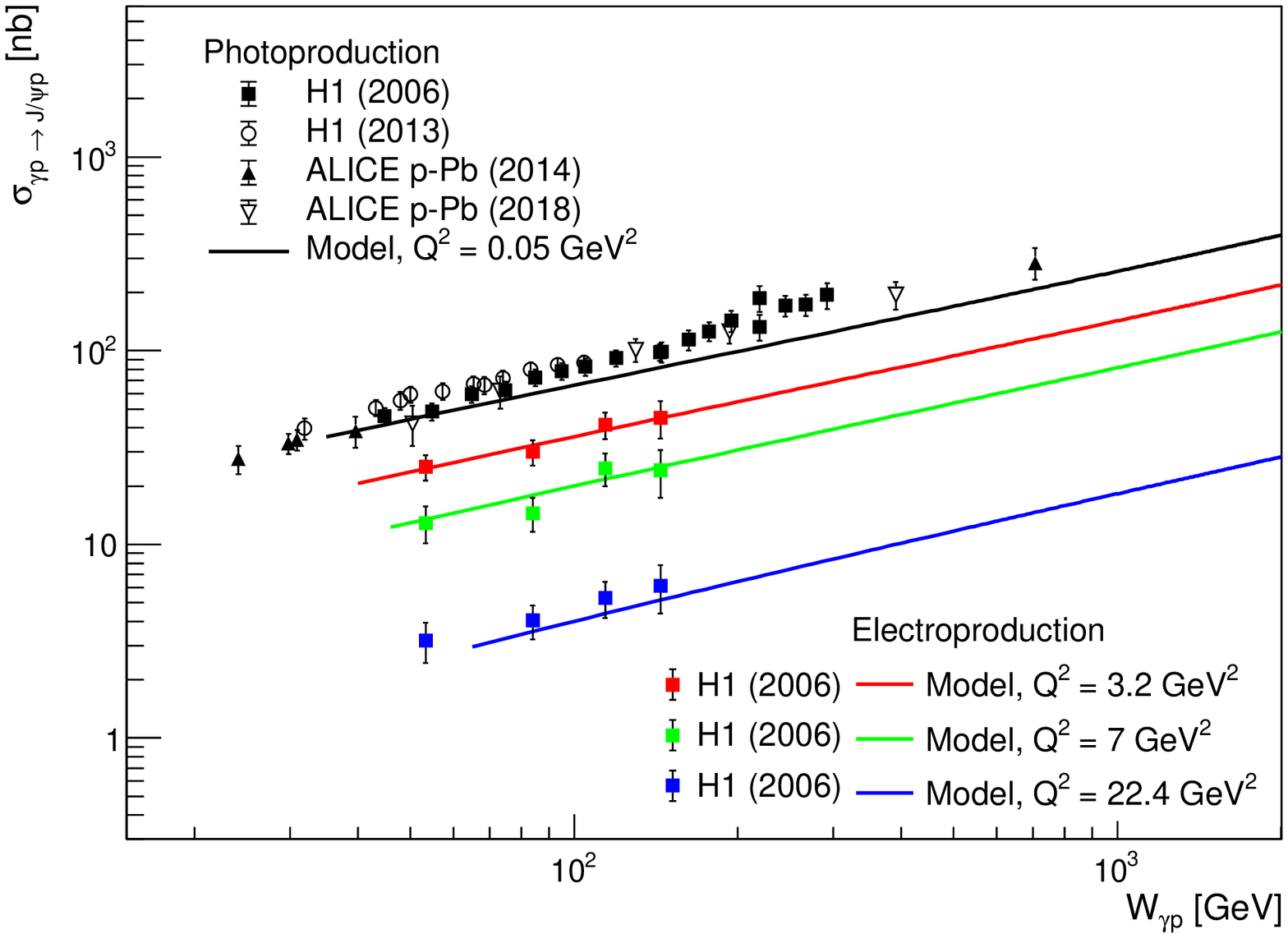}
   \includegraphics[width=0.49\linewidth]{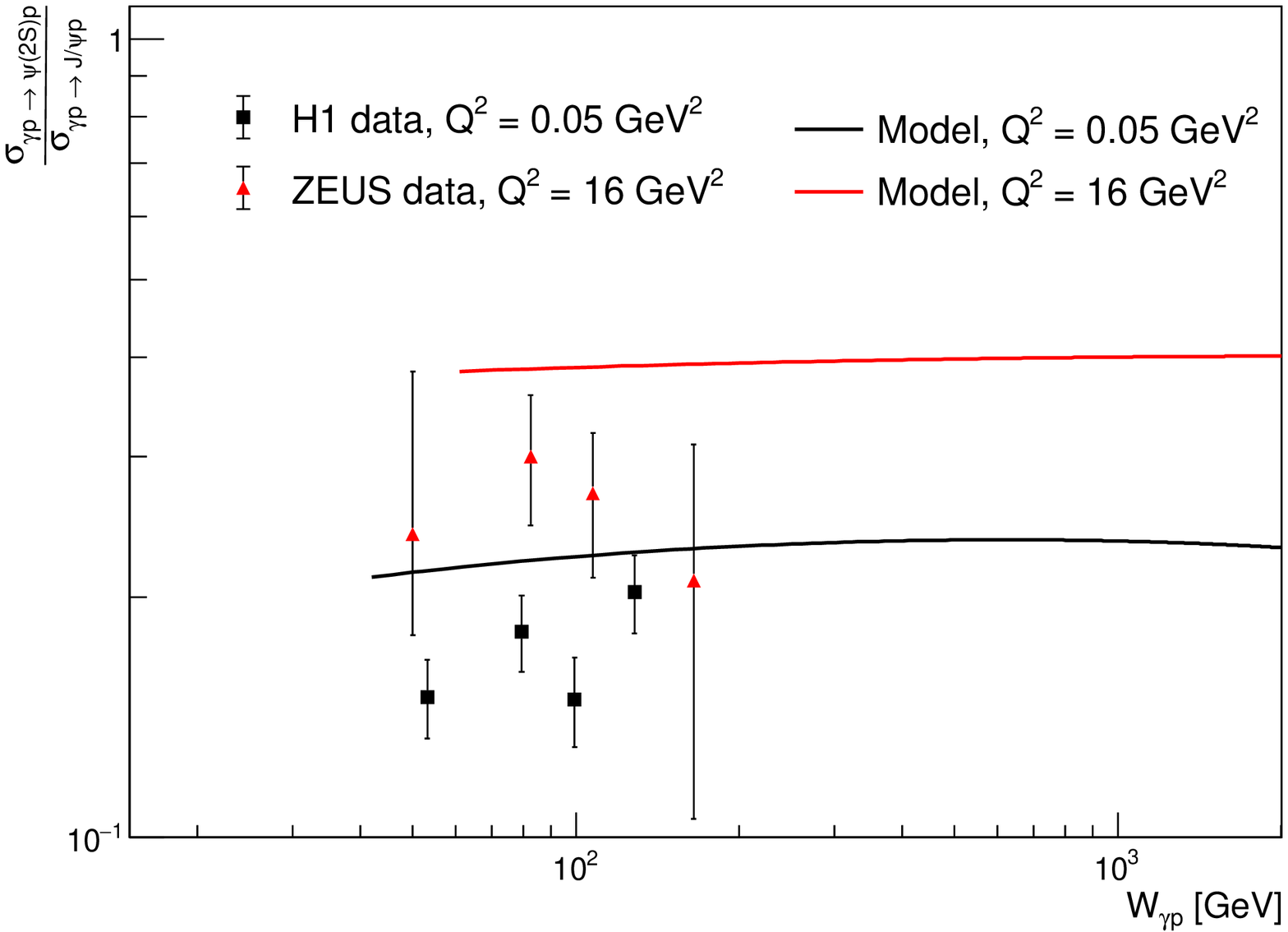}
\caption{Comparison of the predictions of the model (solid lines) with HERA data from H1~\cite{Aktas:2005xu,Alexa:2013xxa} and LHC data from ALICE~\cite{TheALICE:2014dwa,Acharya:2018jua} for the $\Wgp$ dependence of the exclusive photo- and electroproduction cross section of the $\jpsi$ meson (left) and with HERA data from H1~\cite{Adloff:2002re} and ZEUS~\cite{Abramowicz:2016xls} for the $\Wgp$ dependence of the exclusive photo- and electroproduction cross section $\jpsi$/$\psip$ ratio (right).}
\label{VM_total_Jpsi+psi2S}
\end{figure}

The predictions for the exclusive production of the $\jpsi$ meson are compared with the experimental data from H1~\cite{Aktas:2005xu,Alexa:2013xxa} and ALICE~\cite{TheALICE:2014dwa,Acharya:2018jua} experiments in Figs.~\ref{VM_Jpsi_t-distribution}-\ref{VM_total_Jpsi+psi2S}, for several different measurements of kinematic observables. In the left panel of the Fig.~\ref{VM_Jpsi_t-distribution}, the comparison of the $|t|$-distribution of the photoproduction cross section is presented. The predictions give very good agreement with the data at energies $\Wgp = 55 \; \mathrm{GeV}$ and $\Wgp = 100 \; \mathrm{GeV}$. The result for $\Wgp = 78 \; \mathrm{GeV}$ is slightly underestimated at low values of $|t|$, however one can notice the very small difference in the measured  data with respect to the result for $\Wgp = 100 \; \mathrm{GeV}$. Since the value of $\Wgp$ from the experimental data is a mean value estimated from a measured energy range, the result of the model can be considered satisfactory. The same comparison for the electroproduction at three different values of $Q^2$ can be seen in the right panel of the same figure. Although our predictions do not describe all the data points, we conclude the agreement between the data and the model to be qualitatively good. The same conclusion applies to the comparison of the model predictions with the measured $\Wgp$ dependence of the exclusive differential photo- and electroproduction cross sections at several fixed values of $|t|$ presented in Fig.~\ref{VM_Jpsi_fix-t_W-dependence}. The agreement of the predictions with the data is very good at low values of $\Wgp$, however at larger values ($ \sim 10^2 \; \mathrm{GeV}$), the predictions are underestimated when compared to experimental photoproduction data. We have also obtained total cross section for the $\jpsi$ production which is presented in the left panel of the Fig.~\ref{VM_total_Jpsi+psi2S}. The predictions for the electroproduction at three different values of $Q^2$ give a very good description of the available data. The result for photoproduction gives a good agreement with the data at low values of $\Wgp$, however at high energies the result is again underestimated when compared to data.

Also, the exclusive cross section of the $\psip$ meson was calculated within the model. The experimental data are not available for the total cross sections, but only for a ratio of the $\psip$ to $\jpsi$ cross sections, the predictions for these ratios for photoproduction and electroproduction at $Q^2 = 16 \; \mathrm{GeV^2}$ are calculated and compared to data from H1~\cite{Adloff:2002re}, and ZEUS~\cite{Abramowicz:2016xls}, respectively, in the right panel of the Fig.~\ref{VM_total_Jpsi+psi2S}. The description of the data is not very good, yet the large uncertainties of the experimental data do not allow us to make any final conclusions in this case.

\begin{figure}[ht]
  \centering
   \includegraphics[width=0.5\linewidth]{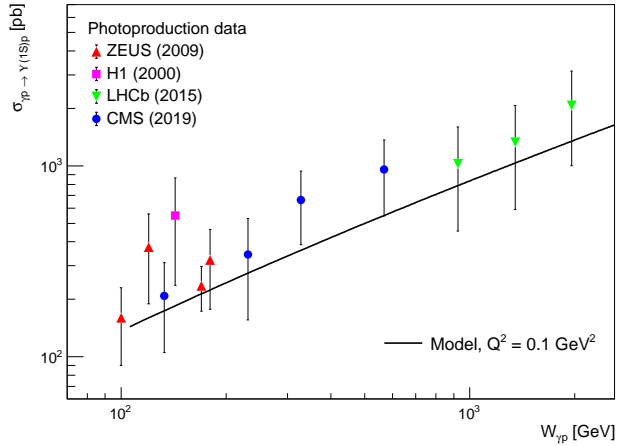}
\caption{Comparison of the predictions of the model (solid lines) with the HERA data from H1~\cite{Adloff:2000vm} and ZEUS~\cite{Chekanov:2009zz}, and LHC data from LHCb~\cite{Aaij:2015kea} and CMS~\cite{Sirunyan:2018sav} for the $\Wgp$ dependence of the exclusive photoproduction cross section of the $\Uos$ meson.}
\label{VM_total_Upsilon}
\end{figure}

To complete the set of the predictions based on the BK equation, the exclusive photoproduction of the $\Uos$ meson is presented in Fig.~\ref{VM_total_Upsilon}. The prediction is compared with experimental data obtained at HERA by H1~\cite{Adloff:2000vm} and ZEUS~\cite{Chekanov:2009zz} experiments. It is also compared with the two latest measurements -- in proton-proton collisions at $\sqrt{s} = 7 \; \mathrm{TeV}$ and $\sqrt{s} = 8 \; \mathrm{TeV}$ by LHCb~\cite{Aaij:2015kea}, and in proton-lead collisions at $\sqrt{s} = 5.02 \; \mathrm{TeV}$ by the CMS experiment~\cite{Sirunyan:2018sav}. The description of the data is good, although the large uncertainties prevent us from making any strong conclusions regarding the agreement of the predictions with the data.

\section{\label{sec:Conclusions}Conclusions}

The solution of the Balitsky-Kovchegov equation with the collinearly improved kernel and including the impact-parameter dependence has been obtained numerically. This solution does not show the so-called Coulomb tails that have appeared in previous attempts to include the impact-parameter dependence. We have shown that the suppression at large values of the impact parameter is due to the suppression of contributions from daughter dipoles of large sizes in the terms of the collinearly improved kernel that deal with the resummation of  double and single collinear logarithms.

The solutions based on a physics-inspired initial condition have been confronted with HERA and LHC data of the structure function of the proton measured in deep-inelastic scattering and of exclusive vector meson photo- and electroproduction.  The predictions described data over a large kinematic range in scale and in energy.

The dipole scattering amplitudes computed in this work are publicly available on the website \url{https://hep.fjfi.cvut.cz/} along with instructions on how to use them.

\section{\label{sec:Acknowledgements}Acknowledgements}
We would like to thank Dionysios Triantafyllopoulos for fruitful discussions regarding this paper. Our work has been partially supported from grant LTC17038 of the INTER-EXCELLENCE program at the Ministry of Education, Youth and Sports of the Czech Republic, by grant 17-04505S of the Czech Science Foundation, GA \v CR and the COST Action CA15213 THOR. Computational resources were provided by the CESNET LM2015042 grant and the CERIT Scientific Cloud LM2015085, provided under the program Projects of Large Research, Development, and Innovations Infrastructures.
\bibliographystyle{apsrev4-1}
\bibliography{references}

\end{document}